\documentclass[10pt,aps,prd,preprint,showkeys,nofootinbib,superscriptaddress]{revtex4-1}
\usepackage{amsmath,amsfonts} 
\usepackage{bbold} 
\usepackage{graphicx,color}
\usepackage{multirow}
\usepackage{geometry}
\usepackage{makecell}
\usepackage[vcentermath]{youngtab}
\usepackage{tikz}
\usepackage{slashed} 
\usepackage{graphics}
\usepackage{graphicx}
\usepackage{setspace}

\usepackage{multirow}
\usepackage{booktabs}    
\usepackage{placeins}
\usepackage{soul}
\usepackage[colorlinks]{hyperref}
\usepackage{chngcntr}
\usepackage{bbm}
\usepackage{amsfonts}
\usepackage{mathrsfs}
\usepackage{amsmath,amssymb,amsbsy,bm}

\newcommand{\mc}[1]{\mathcal{#1}}
\newcommand{\tr}[1]{{\mbox{tr}[#1]}}

\newcommand{\vev}[1]{\langle #1 \rangle}
\newcommand{\ket}[1]{| #1 \rangle}
\newcommand{\bra}[1]{\langle #1 |}

\newcommand{\comments}[1]{}
\newcommand{\eq}[1]{\begin{equation}\begin{split} #1 \end{split}\end{equation}}

\newcommand{\red}[1]{{\color{red}#1}}
\newcommand{\blue}[1]{{\color{blue}#1}}
\begin{document} 
\count\footins = 800

\title{Amplitude/Operator Basis in Chiral Perturbation Theory\vspace{0.2cm}}

\author{Ian Low\vspace{0.1cm}}
\email{ilow@northwestern.edu}
\affiliation{\small{High Energy Physics Division, Argonne National Laboratory, Argonne, IL 60439, USA}}
\affiliation{\mbox{\small{Department of Physics and Astronomy, Northwestern University, Evanston, IL 60208, USA}}}

\author{Jing Shu}
\email{jshu@pku.edu.cn}
\affiliation{\small{School of Physics and State Key Laboratory of Nuclear Physics and Technology, Peking University, Beijing 100871, China}}
\affiliation{\small{Center for High Energy Physics, Peking University, Beijing 100871, China}}

\author{Ming-Lei Xiao}
\email{minglei.xiao@northwestern.edu}
\affiliation{\small{High Energy Physics Division, Argonne National Laboratory, Argonne, IL 60439, USA}}
\affiliation{\mbox{\small{Department of Physics and Astronomy, Northwestern University, Evanston, IL 60208, USA}}}

\author{Yu-Hui Zheng}
\email{zhengyuhui@itp.ac.cn}
\affiliation{\small{CAS Key Laboratory of Theoretical Physics, Institute of Theoretical Physics,Chinese Academy of Sciences, Beijing 100190, China}}
\affiliation{\mbox{\small{School of Physical Sciences, University of Chinese Academy of Sciences, Beijing 100049, China}}}

\begin{abstract}
\vspace{0.3cm}
We establish a systematic construction of the on-shell amplitude/operator basis for Chiral Perturbation Theory (ChPT) in $D=4$ spacetime dimensions and with an arbitrary number of flavors $N_f$. For kinematic factors, we employ spinor-helicity variables to construct the soft blocks, which are local amplitudes satisfying the Adler's zero condition,  as well as to  take into account  the reduction in the kinematic basis due to the Gram determinant, which arises at $O(p^{10})$ when the number of multiplicity $N$ in an amplitude becomes large: $N>D$. For flavor factors,  we include group-theoretic relations at small $N_f$, $N_f\le N$, which decreases the flavor basis. The  result is obtained by adapting the Young tensor method of constructing the operator basis for generic effective field theories to the case of non-linearly realized symmetries. Working in the massless quark limit,  we present purely mesonic operators for both even- and odd-parity at  $O(p^6)$ and $O(p^8)$ for $N=6$ and arbitrary $N_f$, and establish a direct correspondence between the amplitude basis and the operator basis. Furthermore, the redundancy due to the Gram determinant is studied at $O(p^{10})$ for $N=6, 8$ and 10.
\end{abstract}

\maketitle

\tableofcontents
\newpage

\section{Introduction}
Effective field theories (EFTs) are powerful tools to understand salient behaviors of physical systems without detailed knowledge of ultraviolet (UV) physics, which is parameterized by some unknown constants. This remarkable approach, which has existed for half a century, is widely appreciated in several subfields of physics, from condensed matter to high energy and nuclear physics.

The philosophy of constructing an EFT is to use the relevant degrees of freedom to write down all effective operators consistent with symmetries of the physical system under consideration – what is not forbidden must be allowed – which generally results in an infinite number of operators. In order to be predictive, a power counting scheme must be supplied to estimate the relative importance of various effective operators so that only a small number of them contribute to the experimental process of interest.

Moreover, as measurements on a particular system become more and more precise, an increasing number of operators become relevant and need to be incorporated into the EFT. A prime example is the Chiral Perturbation Theory (ChPT) \cite{Weinberg:1968de, Weinberg:1978kz} describing interactions of pions with a single nucleon in a derivative expansion of $\partial/\Lambda$, where $\Lambda\sim O(1\ {\rm GeV})$. At higher orders in $\partial/\Lambda$, the construction of ChPT using the traditional Lagrangian approach becomes quite complicated, and the number of effective operators grows exponentially. In particular, there are subtleties associated with operator redundancies that are inherent in the Lagrangian formulation, rendering the enumeration of a complete and minimal set of operators a formidable task. After enduring efforts over a span of decades, the effective Lagrangian of ChPT is now known up to $O(\partial^8)$ \cite{Gasser:1983yg,Fearing:1994ga,Bijnens:1999sh,Bijnens:2018lez}.

Another example, which has received much attention lately, is the Standard Model Effective Field Theory (SMEFT). Measurements at the Large Hadron Collider (LHC) push the validity of the Standard Model (SM) to unprecedented energy scales beyond the weak scale characterized by the Higgs vacuum expectation value $v=246$ GeV. As a consequence, any heavy, beyond-the-SM effects can be parameterized by irrelevant operators with increasing mass dimensions. It turned out there is only a single operator at the dim-5 order \cite{Weinberg:1979sa} that is consistent with the symmetries of the SM. At the dim-6 order \cite{Buchmuller:1985jz,Grzadkowski:2010es}, the construction quickly becomes laborious, as in the case of the ChPT. Moreover, operator redundancies become constant sources of confusion and debate, especially when one attempts to go either to higher orders in the mass dimensions or beyond the approximate flavor symmetries of the SM. 

The complexity of these EFTs at higher orders in the respective power counting schemes highlights the limitation of the traditional formulations and calls for new approaches to the problem. In the past few years, we have seen tremendous progress in applying new perspectives and methodology to constructing both ChPT and SMEFT. For the ChPT, which involves pseudo-Nambu-Goldstone bosons (NGB's) and nonlinearly realized symmetries, a new infrared (IR) perspective based on the single soft theorem of NGB's, i.e., the Adler's zero \cite{Adler:1964um}, was developed both in the Lagrangian formalism \cite{Low:2014nga,Low:2014oga} and in the on-shell formalism \cite{Cheung:2014dqa,Cheung:2015ota}. The IR perspective facilitated assembling an operator basis, in terms of the kinematic invariants $s_{ij}=2p_i\cdot p_j$,  for independent operators up to $O(p^{10})$ in Ref.~\cite{Low:2019ynd,Dai:2020cpk}.\footnote{\setstretch{1} In addition to ChPT, the IR perspective also has important implications for extensions of the SM where the Higgs arises as a pseudo-Nambu-Goldstone boson\cite{Liu:2018vel,Liu:2018qtb}.} (If one is only interested in enumerating the number of operators, there is also the Hilbert series method in Ref.~\cite{Graf:2020yxt}.) In the SMEFT, advances were made in searching for an independent set of operator basis beyond the dim-6 order \cite{Henning:2015daa,Li:2020gnx,Li:2020xlh,Li:2020xlh,Li:2020zfq,Graf:2020yxt,Li:2020tsi,Murphy:2020rsh}. In particular, a standard operator basis called y-basis (y stands for Young tableau) was introduced in Refs.~\cite{Li:2020gnx,Li:2020xlh,Li:2020tsi,Li:2020zfq} and a \texttt{Mathematica} package ABC4EFT was made public on \href{https://abc4eft.hepforge.org/}{HEPForge} \cite{Li:2022tec}. 

There are, however, limitations to these new developments. For example, the studies in Refs.~\cite{Bijnens:2018lez,Dai:2020cpk} were done without considering the kinematic constraint from the vanishing Gram determinant, which arises from the fact that in a spacetime dimension $D$, any $N>D$ momenta must be linearly dependent.\footnote{\setstretch{1} More specifically, linear dependence of any $N>D$ vectors $X_i^\mu$ in $D$ dimensional spacetime implies vanishing of the Gram determinant $\det ({X_i\cdot X_j}) = 0,  \ \ i,j=1,\cdots, N$.}  These works further assumed a large number of flavor $N_f$, ignoring group-theoretical relations on the flavor factors which may exist at a small $N_f$.  These issues will further reduce the number of independent operators in ChPT, where $D=4$ and $N_f=2,3$. Note that the vanishing Gram determinant is automatically fulfilled in on-shell spinor-helicity variables employed in the method presented in Refs.~\cite{Li:2020gnx,Li:2020xlh,Li:2020tsi,Li:2020zfq}, which nonetheless cannot be applied directly to EFTs with nonlinearly realized symmetries such as the ChPT. In this work, we will fill in such a gap and present an algorithm, modified from Refs.~\cite{Li:2020gnx,Li:2020xlh,Li:2020tsi,Li:2020zfq},  to systematically obtain an amplitude/operator basis for ChPT to arbitrary orders in the momentum expansion. Similar to Ref.~\cite{Dai:2020cpk} we work in the limit of massless quarks and focus on purely mesonic operators without external sources.

More specifically, in this work we construct an operator basis using on-shell variables with the following features:
\begin{itemize}
    \item Adler's zero condition is satisfied.
        
    \item Vanishing Gram determinants at $D=4$.              
		
	\item Parity-odd operators as well as the parity-even ones.
		
	\item Reduction of the soft basis for finite $N_f$.
\end{itemize}
It is also shown that with the similar technique of y-basis reduction, the operator basis for ChPT with external sources of fermions or vectors can also be obtained \cite{Sun:2022ssa}.

The paper will be organized as follows:
In section 2, we present the general features of the Young Tableau basis (y-basis) of the Lorentz structure and show how it solves the subtleties in the literature and generates the soft blocks. Section 3 is dedicated to the flavor structures, where we explicitly show the redundancies of trace structures at small $N_f$. We combine the kinematics and flavor structures in section 4 by introducing the trace orbit, and show how the Gram determinant redundancy is taken into account in our algorithm. An explicit example of obtaining the $O(p^6)$ 6-point soft basis is provided in section 5. Finally, we summarize in section 6. We provide some background on group theory and the $O(p^8)$ 6-point soft basis in the appendices.


\section{The Kinematic Factor}

The starting point of our construction is the "soft block" introduced in Ref.~\cite{Low:2019ynd} as the seed for the "soft recursion" relation for Nambu-Goldstone bosons \cite{Cheung:2014dqa}. Soft blocks are contact interactions that satisfy the Adler's zero condition \cite{Adler:1964um} when all external legs are on-shell. They are in one-to-one correspondence with the independent operators in momentum expansion in ChPT \cite{Low:2019ynd}:
 \eq{
    \text{Soft block }\mc{B}_i \quad \Leftrightarrow \quad \text{Invariant operators }\mc{O}_i \ .
}
The Lorentz part of the soft block is a function of the on-shell momenta $p_i$ of the NGBs, which corresponds to the operator as
\eq{\label{eq:block_correspond}
    p_{i\mu_1}p_{i\mu_2}\cdots p_{i\mu_n}\quad \Leftrightarrow\quad \nabla_{(\mu_1} \nabla_{\mu_2} \cdots \nabla_{\mu_{n-1}} u_{\mu_n)},
}
where $u_\mu \equiv e^{-i\pi}\nabla_\mu e^{i\pi}$ is the field operator in the CCWZ formalism that generates an NGB $\bra{\pi}u_\mu\ket{0} \sim p_\mu$ and is covariant under the leftover symmetry $H$. The Lorentz indices in the corresponding operator are totally symmetrized. Building blocks with indices not totally symmetrized can always be cast into the symmetrized one with extra pieces proportional to the so-called Equation of Motion (EOM) redundancy \cite{Li:2020zfq}.

We will use ``soft basis'' to refer to an independent basis for the soft blocks, which can also be thought of as the leading contact on-shell interactions from each invariant operator in ChPT. Enumerating and constructing the soft blocks then gives an on-shell construction of ChPT via soft recursion relation. Ref.~\cite{Dai:2020cpk} listed the soft blocks for purely mesonic operators in the ChPT up to $O(p^{10})$ using momentum invariants $(p_i+p_j)^2=2p_i\cdot p_j$, which did not take into account constraints from the Gram determinant.
The Adler zero condition requires that the $N$-point soft blocks should satisfy the single soft limits of each NGB
\eq{\label{eq:adler_zero}
    \lim_{p_i\to0}\mc{B}(p_1,\dots,p_i,\dots,p_N) = 0, \qquad \forall\ i = 1,\dots,N.
}
Hence we transform the question of obtaining an operator basis for ChPT into a problem of constructing local, contact amplitudes that satisfy eq.~\eqref{eq:adler_zero}.

In general the amplitudes can be written as the product of a kinematic part $\mc{M}(p_1,\dots,p_N)$ and a flavor structure part $\mc{T}^{a_1,\dots,a_N}$, where $a_i$ is the flavor index of the NGB under the unbroken global symmetry $H$. In  ChPT, $a_i$ are adjoint indices of $SU(N_f)$ group, and the invariant tensors $\mc{T}^{a_1,\dots,a_N}$ can always be written as combinations of trace structures, which form a linear space with an independent basis $\{\mc{T}_i\}$. The kinematic part also forms a linear space with basis $\{\mc{M}_i\}$, all of which satisfy the Adler's zero condition. Hence each soft block  $\mc{B}_i$ can be expressed as the following linear combination:
\eq{
    \mc{B}_i = \sum_{j,k}\ C_i^{jk}\, \mc{M}_j(p_1,\dots,p_N)\ \mc{T}_k^{a_1,\dots,a_N},
}
where $C_i^{jk}$'s are some constants. The soft block must also satisfy the Bose symmetry and be invariant under permutations of external legs:
\eq{
    \sigma\circ \mc{B}_i \equiv \sum_{j,k}\ C_i^{jk}\,\mc{M}_j(p_{\sigma(1)},\dots,p_{\sigma(N)})\ \mc{T}_k^{a_{\sigma(1)},\dots,a_{\sigma(N)}} = \mc{B}_i\ ,
}
where $\sigma\in S_N$ is an element of the permutation group.

In this section, we focus on obtaining a basis of kinematic factors $\mc{M}_i$. For purely NGB amplitudes, $\mc{M}_i$ are polynomials in the   Lorentz invariant building blocks 
\eq{
s_{ij} \equiv g^{\mu\nu}p_{i\mu}p_{j\nu}\quad \text{and} \quad \epsilon(i,j,k,l) \equiv \epsilon^{\mu\nu\rho\sigma}p_{i\mu}p_{j\nu}p_{k\rho}p_{l\sigma}\,.
} 
However,  polynomials in these building blocks are not all independent. The most important relations are massless on-shell condition $p_i^2=0$ and momentum conservation, which reads $\sum_ip_i=0$ in a convention where all momenta are incoming. They reduce, for instance, the set of independent kinematic invariants for 6-point (pt) amplitudes to the set 
\eq{\label{eq:Mandelstam_6}
    \{s_{24}, s_{25}, s_{26}, s_{34}, s_{35}, s_{36}, s_{45}, s_{46}, s_{56}\}\ ,
}
which obviously is not a unique choice. In general, for $N$-pt amplitudes with $N\le D+1$ the on-shell constraint and momentum conservation gives rise to $N$ linearly independent constraints:
\eq{
\sum_{j=1,j\neq i}^N s_{ij} = 0 \ ,\qquad \forall\ \ i = 1,\dots,N\ .
}
These reduce the dimensionality of the kinematic space to ${N \choose 2} - N =N(N-3)/2$.

For $N> D+1$ there are additional relations, such as the CP even Gram determinant 
\eq{\label{eq:D4_even}
\det (s_{ij}) = 0\ ,\qquad \forall \ \ i,j=1,2,\dots,D+1\,,
}
and the CP odd combination
\eq{\label{eq:D4_odd}
\sum_{\sigma\in S_{D+1}}{\rm sgn}(\sigma)\times
s_{i,\sigma(j)}\epsilon(\underbrace{\sigma(k),\sigma(l),\dots}_{D}) = 0
}
both due to the fact that $D+1$ momentum vectors in $D$ dimensional spacetime cannot be anti-symmetrized.
In particular, the Gram determinant itself is a polynomial of degree $(D+1)$ in $s_{ij}$, which would reduce the number of independent polynomials of degree $D+1$ and more. Therefore the constraint from the Gram determinant becomes effective at the order of $p^{2(D+1)}$ or higher.\footnote{\setstretch{1} In $D=4$ the Gram determinant only takes effect at the level of degree-5 polynomials in $s_{ij}$, or at $p^{10}$ order, as we will show explicitly later. 
} Specializing to $D=4$ from now on, we will employ the spinor helicity variables $(\lambda,\tilde\lambda)$ to construct the kinematic factors $\mc{M}_j(p_{\sigma(1)},\dots,p_{\sigma(N)})$. Recall that in $D=4$ a massless momentum $p_\mu$ satisfies the on-shell condition $p^2=p_\mu p^\mu = {\rm det}(p\cdot \sigma)= 0$, where $\sigma^\mu =(1, \vec{\sigma})$. In turn this implies $p\cdot \sigma$ has rank-1 and can be written as the outer product of two two-component spinors $(\lambda,\tilde\lambda)$: $p_{\alpha\dot{\alpha}}\equiv (p\cdot \sigma)_{\alpha\dot{\alpha}}=\lambda_\alpha\tilde\lambda_{\dot{\alpha}}$.\footnote{\setstretch{1} For real momentum $p\cdot \sigma$ is Hermitian and $\lambda^\dagger = \pm \tilde\lambda.$} The spinor helicity variables are intrinsically $D=4$ and take into account the constraint from the Gram determinant by default through  the Schouten identities \cite{Elvang:2013cua}:
\eq{\label{eq:schouten}
    & \vev{ij}\lambda_k + \vev{jk}\lambda_i + \vev{ki}\lambda_j = 0, \quad \vev{ij} \equiv \lambda_i^\alpha\lambda_{j\alpha}, \\
    & [ij]\tilde\lambda_k + [jk]\tilde\lambda_i + [ki]\tilde\lambda_j = 0, \quad [ij] \equiv \tilde\lambda_{i\dot\alpha}\tilde\lambda_i^{\dot\alpha} \ ,
}
where the angle $\vev{\cdot }$ and square brackets $[\cdot]$ are defined as
\eq{\label{eq:angsq}
	& \vev{ij} \equiv \epsilon_{\alpha\beta}\lambda_{i}^{\alpha}\lambda_{j}^{\beta} ,\qquad 
	[ij] \equiv \tilde\epsilon^{\dot\alpha\dot\beta}\tilde\lambda_i{}_{\dot\alpha}\tilde\lambda_j{}_{\dot\beta}\ .
}
Moreover, using trace identities for Pauli matrices one can show
\eq{\label{eq:sijanbr}
s_{ij} \equiv (p_i+p_j)^2= 2p_i\cdot p_j = [ij]\vev{ji} \ .
}
Another advantage of adopting spinor variables is that external sources can be introduced in a consistent way \cite{Henning:2019enq,Li:2020gnx,Li:2020xlh}. Last but not least, we have to impose the Adler's zero condition.

\subsection{Overview on the Reduction of Lorentz Structures}

Let us first present the basic ingredients for getting an independent basis of Lorentz structures in a general quantum field theory. (For details please refer to Refs.~\cite{Li:2020gnx,Li:2020xlh}). The starting point is to classify all matter content according to  irreducible representations of the Lorentz group $SL(2,C)\approx SU(2)_l \times SU(2)_r$ and use the notation $\phi \in (0,0)$, $\psi \in (\frac12,0)$, $\psi^\dagger \in (0,\frac12)$, $F_L \in (1,0)$, $F_R \in (0,1)$ and etc, where we are using a chiral basis for the gauge field $F_{L/R}=\frac12(F\pm i\tilde{F})$.  Moreover, a covariant derivative $D \in (\frac12,\frac12)$ representation. An operator carrying $k$ covariant derivatives is then represented by an on-shell local amplitude specified by external states with definitive helicities $\{h_1,h_2,\dots,h_N\}$ and the power of momenta $k$.   The numbers of angle brackets $n$ and square brackets $\tilde{n}$ in the local amplitude that corresponds to the effective operator are given by
\eq{\label{eq:nntilde}
	n = \frac{k}{2} - \sum_{h_i<0}h_i ,\qquad \tilde{n} = \frac{k}{2} +\sum_{h_i>0}h_i \ ,
}
which is easy to understand from  the Little Group (LG) scalings $e^{ih_i\varphi_i}$ of the particles (and derivatives) involved as well as the scaling rules for the spinors $\lambda_i \to e^{-i\varphi_i/2}\lambda_i$, $\tilde\lambda_i \to e^{i\varphi_i/2}\tilde\lambda_i$.

As an example, consider an operator of the type $\psi\psi^\dagger\phi F_L D$. It has $(\tilde{n},n)=(1,2)$ according to eq.~(\ref{eq:nntilde}), and the local amplitude with the correct LG property is 
\eq{\label{eq:amp_example}
	\mc{M}^{d=7}(\psi_1,\psi^\dagger_2,\phi_3,F_{L,4}) = \vev{14}[23]\vev{34}.
}
According to Refs.~\cite{Henning:2019enq,Li:2020gnx}, the amplitude eq~\eqref{eq:amp_example} can be represented by a Young tableau,
\eq{\label{eq:amp-yt_example}
	\vev{14}[23]\vev{34}  = \mathcal{E}^{14ij}[ij] \times \vev{14}\vev{34} \simeq \young({{\color{blue}1}}13,{{\color{blue}4}}44)\ ,
}
blue columns represent square brackets $[\cdot]$ (up to a Levi-Civita tensor $\mathcal{E}$) while black columns stand for angle brackets $\vev{\cdot}$. The detailed definition is given in the appendix~\ref{app:fock}.
The above Young tableau is a specific state in the $\tiny\yng(3,3)$, indicating a so-called \emph{primary} representation space of class $(\tilde{n},n)=(1,2)$, of the $SU(N)$ group, where $N=4$ is the number of particles involved, and $\mathcal{E}$ is the Levi-Civita tensor of the $SU(N)$ group, under which  $(\lambda_1, \cdots, \lambda_N)$ or $(\tilde{\lambda}_1, \cdots, \tilde{\lambda}_N)$ transforms as the (anti-)fundamental representation.

The \emph{primary} representation space is defined as that annihilated by the special conformal generator $K$ \cite{Henning:2019enq} which, by analogy with the Laplacian $\nabla^2$ acting on spatial functions restricted at fixed radius $r$ giving solutions as harmonic functions, generates the \emph{harmonic} spinor functions as an independent basis for fixed $P\equiv \sum_i p_i=0$.\footnote{\setstretch{1} We use all-incoming convention for the momenta $p_i$, i.e. physical out-going momenta are expressed as $-p$ with $p^0<0$.} 
It turns out that the primary representations are given by Young Diagrams (YD) with $n$ 2-row columns and $\tilde{n}$ $(N-2)$-row columns, thus we have a $2\times3$ YD for the $(\tilde{n},n)=(1,2)$-type amplitudes as shown in eq.~\eqref{eq:amp-yt_example}.
Group theory asserts that an independent basis of an irreducible representation space can be obtained as the Semi-Standard Young Tableau (SSYT)\cite{ma2007group}, which we recall are YD filled with numbers from 1 to $N$ subject to the rules that i) the numbers in a row increase weakly from left to right and ii) the numbers in a column increase strongly from top to bottom. All the other Young tableaux could be expressed as linear combinations of SSYTs through the Fock conditions, which are equivalent to the Schouten Identities in eq.~(\ref{eq:schouten}) and total momentum conservation.\footnote{\setstretch{1} Please refer to the appendix~\ref{app:fock} for details.}
For instance, there is an equivalent amplitude $\vev{14}^2[12]$ to eq.~(\ref{eq:amp_example}) which also satisfies all the LG scaling and corresponds to a non-SSYT. Using Fock conditions it can be expressed in terms of SSYT as
\eq{
	\vev{14}^2[12] \simeq \young({{\color{blue}3}}11,{{\color{blue}4}}44) = \young({{\color{blue}1}}13,{{\color{blue}4}}44) \simeq \vev{14}[23]\vev{34}.
}
Using momentum conservation one can see explicitly,
\eq{
	\vev{14}^2[12] = -\vev{14}[2|p_1\ket{4} = \vev{14}[2|(p_2+p_3+p_4)\ket{4} = \vev{14}[23]\vev{34}.
}
The isomorphism between the Fock conditions and the amplitude equivalence, including momentum conservation and the Schouten identities, guarantees that the SSYT forms an independent basis of the contact amplitudes.

There are often a lot of SSYTs for a given YD, associated with various types of operators, some even with non-physical particles $|h|>2$. We observe that to find the particular subset of SSYTs that correspond to a certain external state, it is sufficient to only look at a specific collection of labels to fill in.  The general rule is that $\# i$, defined as the frequency the particle label $i$ would appear in the SSYT subset, is given by
\eq{\label{eq:fill}
	\# i = \tilde{n}-2h_i.
}
Going back to the type $\psi_1\psi^\dagger_2\phi_3 F_{L4} D$, which has $(n,\tilde{n})=(2,1)$, we have in this case $(\#1,\#2,\#3,\#4)=(2,0,1,3)$ and the collection of particle labels to be used to fill in the YD is $\{1,1,3,4,4,4\}$. Therefore, the independent basis of amplitudes for a given type of operator, which we call the \emph{y-basis}, consists of all the SSYTs with the various ordering of the same collection of labels.  
The number of such SSYT $\mathbb{d}_{\mc{M}}$, which is exactly the number of independent Lorentz structures for the given state, can be computed by the tensor product decomposition\footnote{The tensor product of Young diagrams can be derived by the Littlewood-Richardson rules.} (TPD)
\eq{\label{eq:SSYT_count}
	\bigotimes_{i=1}^{N} \underbrace{\young(ii{{\dots}}i)}_{\# i} =  \text{SSYD}_1 \oplus \dots \oplus \text{SSYD}_{\mathbb{d}_{\mc{M}}} \oplus \text{non-SSYD's}
}
where ${\scriptsize\young(ii{{\dots}}i)}$ is the contribution of the $i$th particle. 
For $\psi_1\psi^\dagger_2\phi_3 F_{L4} D$, we have
\eq{
    \young(11)\otimes\young(3)\otimes\young(444) = \underbrace{\young({{\blue{1}}}13,{{\blue{4}}}44)}_{\text{SSYD}}
    \oplus\ \underbrace{ \young({{\blue{1}}}14,{{\blue{3}}}4,{{\blue{4}}})}_{\text{non SSYD}}\oplus\dots
}
where only one SSYT is obtained. Thus eq.~\eqref{eq:amp_example} is the only independent local amplitude.

Another important technique is to express any given amplitude in terms of linear combinations of the y-basis \cite{Li:2020xlh,Li:2022tec}.
The procedure involves a particular order to apply the Fock conditions to the corresponding Young tableau, or equivalently to apply the momentum conservation and Schouten identities to the amplitude, which are elaborated in the appendix~\ref{append:gram}.
To have a taste of how the reduction works, we take an example relevant to the ChPT to be discussed. 
The $n(n-3)/2$ independent Mandelstam variables can be derived as SSYT of the type $\phi^n D^2$ operators, which has $(n, \tilde{n})=(1,1)$ according to eq.~(\ref{eq:nntilde}). Moreover, each particle label appear only once, as $\#i=1$ by the fill rule in eq.~(\ref{eq:fill}). The 9 independent kinematic invariants in eq.~\eqref{eq:Mandelstam_6} then correspond to the 9 SSYTs:
\eq{
	\begin{array}{ccccccccc}
		\young({{\color{blue}1}}2,{{\color{blue}3}}4,{{\color{blue}5}},{{\color{blue}6}}) \quad  & \quad \young({{\color{blue}1}}2,{{\color{blue}3}}5,{{\color{blue}4}},{{\color{blue}6}}) \quad  & \quad \young({{\color{blue}1}}2,{{\color{blue}3}}6,{{\color{blue}4}},{{\color{blue}5}}) \quad  & \quad \young({{\color{blue}1}}3,{{\color{blue}2}}4,{{\color{blue}5}},{{\color{blue}6}}) \quad  & \quad \young({{\color{blue}1}}3,{{\color{blue}2}}5,{{\color{blue}4}},{{\color{blue}6}}) \quad  & \quad \young({{\color{blue}1}}3,{{\color{blue}2}}6,{{\color{blue}4}},{{\color{blue}5}}) \quad  & \quad \young({{\color{blue}1}}4,{{\color{blue}2}}5,{{\color{blue}3}},{{\color{blue}6}}) \quad  & \quad \young({{\color{blue}1}}4,{{\color{blue}2}}6,{{\color{blue}3}},{{\color{blue}5}}) \quad  & \quad \young({{\color{blue}1}}5,{{\color{blue}2}}6,{{\color{blue}3}},{{\color{blue}4}})
	\end{array} \ .
} 
It is then straightforward to translate these SSYDs into the corresponding kinematic invariants using the previously stated rules that blue columns represent $[\cdot]$, up to the $SU(N)$ Levi-Civita tensor ${\cal E}$, and black columns denote $\vev{\cdot}$. For instance, using eq.~(\ref{eq:sijanbr}),
\eq{
	\young({{\color{blue}1}}2,{{\color{blue}3}}4,{{\color{blue}5}},{{\color{blue}6}})
	\simeq-[24]\vev{24}=s_{24}\ ,
} 
and so on and so forth.		

To build higher-order amplitudes involving more momenta, instead of generating monomials from these building blocks, we build them by directly constructing SSYTs of larger YD, which automatically include the odd-parity amplitudes. 
Not all the monomials of Mandelstam variables are in the y-basis, as some of them may correspond to non-SSYT. Take for example the $O(p^4)$ amplitude $s_{26}s_{45}$ for type $\phi^6D^4$, whose corresponding Young tableau is not SSYT. We can reduce it to the y-basis as follows:
\eq{
	\begin{array}{c|ccc}
	\hline\\
	\text{Amplitude} & s_{26}s_{45}	& \stackrel{\text{reduce}}{\Rightarrow} &	s_{24}s_{56}+s_{25}s_{46}-(\vev{24}[46]\vev{65}[52]+c.c.) \\[0.2cm]
	\hline\\
	\text{Young tableau} & \young({{\color{blue}1}}{{\color{blue}1}}24,{{\color{blue}3}}{{\color{blue}2}}65,{{\color{blue}4}}{{\color{blue}3}},{{\color{blue}5}}{{\color{blue}6}}) & \stackrel{\text{Fock condition}}{\Rightarrow} &
	\young({{\color{blue}1}}{{\color{blue}1}}25,{{\color{blue}2}}{{\color{blue}3}}46,{{\color{blue}3}}{{\color{blue}5}},{{\color{blue}4}}{{\color{blue}6}}) +\young({{\color{blue}1}}{{\color{blue}1}}24,{{\color{blue}2}}{{\color{blue}3}}56,{{\color{blue}3}}{{\color{blue}4}},{{\color{blue}5}}{{\color{blue}6}}) -\young({{\color{blue}1}}{{\color{blue}1}}25,{{\color{blue}2}}{{\color{blue}3}}46,{{\color{blue}3}}{{\color{blue}4}},{{\color{blue}5}}{{\color{blue}6}}) -\young({{\color{blue}1}}{{\color{blue}1}}24,{{\color{blue}2}}{{\color{blue}3}}56,{{\color{blue}3}}{{\color{blue}5}},{{\color{blue}4}}{{\color{blue}6}}) \\[1cm]
	\hline
	\end{array}\notag
}
While at this order, the y-basis is just an alternative of the Mandelstam monomials, it becomes powerful at orders $\phi^6 D^{n\ge10}$ where the redundancy of the Gram determinant comes into effect. When re-writing the Gram determinant constraints in terms of the spinor-helicity variables, which are inherent $D=4$ construction, one can show that eqs.~\eqref{eq:D4_even} and \eqref{eq:D4_odd}  vanish identically when repeatedly applying the reduction rules described in appendix \ref{append:gram}. 

As an example, consider the monomials of degrees 5  in the 9 kinematic invariants for 6-pt amplitudes in eq.~\eqref{eq:Mandelstam_6}:
\eq{s_{24}^{x_1}s_{25}^{x_2}\dots s_{56}^{x^9}\ , \qquad x_i\ge0  \quad \text{and} \quad \sum_i x_i = 5 \ .}
Without Gram determinant one would get $
{9+5-1 \choose 5} =1287$ basis amplitudes. On the other hand in the y-basis we obtain 1286 independent parity-even amplitudes and the difference here is precisely due to the Gram determinant redundancy. In Table \ref{tab:Gram} we list the number of independent parity-even and parity-odd local amplitudes (kinematic invariants) in general $D$ (without Gram determinant constraints) and in $D=4$, up to $N=8$ and $O(p^{10})$. The reduction in the amplitude basis is significant as we go to higher orders in the momentum expansion, especially for parity-odd amplitudes.

\begin{table}[tbp]
\centering
\begin{tabular}{l|c|c|c|c|c|c|c|c|c|c|c|c}
    \hline\hline 
    &$\phi^6p^6$  & $\phi^6p^8$ & $\phi^6p^{10}$ & $\phi^6p^{12}$ & $\phi^6p^{14}$ & $\phi^7p^6$ &  $\phi^7p^8$ & $\phi^7p^{10}$ & $\phi^7p^{12}$ & $\phi^7p^{14}$ & $\phi^8p^{10}$ & $\phi^8p^{12}$\\
    \hline
    {\small P-even}, {\small General $D$} & - & - & 1287 & 3003 & 6435 & - & - & 8568 & 27132 & 77520 & 42504 & 177100 \\
    {\small P-even}, {\small $D=4$}  & - & - & 1286 & 2994 & 6390 & - & - & 8547 & 26873 & 75790 & 42308 & 173915\\ 
    \hline\hline 
    {\small P-odd},  {\small  General $D$} & 45 & 225 & 825 & 2475 & 6435 & 210 & 1575 & 8400 & 35700 & 128520 & 53900 & 309925 \\
    {\small P-odd},  {\small $D=4$} & 40 & 180 & 600 & 1650 & 3960 & 175 & 1106 & 5019 & 18305 & 56980 & 28196 & 132335 \\
    \hline\hline
\end{tabular}
\caption{\em The number of independent local amplitudes, or kinematic invariants, in both general $D$ and $D=4$. The difference is due to the constraints in eqs.~\ref{eq:D4_even} and \ref{eq:D4_odd}. For parity-even operators eq.~\ref{eq:D4_even} is effective only at the order $p^{n\ge 2D}$. The numbers are computed using the package ABC4EFT \cite{Li:2022tec}}\label{tab:Gram}
\end{table}

\subsection{ Adler's Zero Condition and the Soft Blocks}

To build an operator/amplitude basis for ChPT, the next step is to find all the combinations of the y-basis which satisfy the Adler's zero conditions. These are the soft blocks and  correspond to the leading contact interactions of each invariant operator in ChPT, while the rest of the y-basis may come from the sub-leading contributions and are not described by independent low energy coefficients due to the constraint of non-linear symmetry. 

Let's consider a 6-pt soft block corresponding to the invariant operator $\phi^6D^6$. 
The characteristics of the state is $(n,\tilde{n}) = (3,3)$, hence the primary YD is
$$ \yng(6,6,3,3) \ , $$
while the collection of labels to fill in are $\{1,1,1,2,2,2,3,3,3,4,4,4,5,5,5,6,6,6\}$ due to eq.~\eqref{eq:fill}. With the \texttt{Mathematica} package ABC4EFT, the SSYTs can be enumerated instantaneously with the total number of $\mathbb{d}_{\mc{M}}^{6,6} = 205$, one instance of which reads:
$$ \mc{M}^{\rm y}_{95} = \begin{array}{c} \young({{\color{blue}1}}{{\color{blue}1}}{{\color{blue}1}}235,{{\color{blue}2}}{{\color{blue}2}}{{\color{blue}4}}566,{{\color{blue}3}}{{\color{blue}3}}{{\color{blue}5}},{{\color{blue}4}}{{\color{blue}4}}{{\color{blue}6}}) \end{array} \sim [56]^2[23]\vev{25}\vev{36}\vev{56} .$$
To find an independent subspace that satisfies the Adler zero condition, we first provide a general algorithm that could potentially be applied to NGB amplitudes involving external sources. For purely mesonic operators, we give a second and more direct construction.

The general algorithm is similar to that in \cite{Dai:2020cpk}. We start with a generic combination
\eq{
	\mc{M}^{6,6} = \sum_{i=1}^{\mathbb{d}_{\mc{M}}^{6,6}}\ c_i\, \mc{M}^{\rm y}_i,
}
and impose the Adler's zero condition for $\phi_3$ by taking the soft limit $p_3 \to 0$ and requiring $\mc{M}^{6,6}=0$. In the spinor-helicity formalism the soft momentum can be achieved by the holomorphic soft limit, $|3]\to \alpha |3]$, $\alpha\to 0$.\footnote{Or, equivalently, the anti-holomorphic soft limit $|3\rangle \to \alpha |3\rangle$, $\alpha\to 0$. For a scalar theory the holomorphic and anti-holomorphic soft limits are equivalent, as $\ket{i}$ and $|i]$  appear simultaneously for scalar $i$.} 
This way, we get $n_3=149$ out of 205 terms that are non-zero, but it is not enough to deduce $c_i=0$ for all the 149 terms, because the $\mc{M}^{\rm y}_i$, though independent as 6-scalar amplitudes, are not independent any more after imposing $p_3=0$. We treat them as 5-scalar amplitude instead, and use the $O(\phi^5 p^6)$ order y-basis $\mc{M}^{\rm y(5)}_j$ of the particles $\{1,2,4,5,6\}$ (which have 40 independent SSYTs) to reduce the 149 amplitudes,
\eq{
    \mc{M}^{\rm y}_i\ \stackrel{p_3\to0}{\longrightarrow}\ \sum_{j^\prime=1}^{40}\ \mc{K}_{3,ij^\prime}\,\mc{M}^{\rm y(5)}_{j^\prime},
}
resulting in 149 combinations of the 5-scalar y-basis. Now that $\mc{M}^{\rm y(5)}_{j^\prime}$ are indeed independent amplitudes, their coefficients should vanish due to the constraint, therefore
\eq{
	\sum_{i=1}^{205}\ c_i\, \mc{K}_{3,ij^\prime} = 0 \ .
}
These are the constraints from imposing the Adler's zero condition on particle 3. The full constraints are obtained by imposing the Adler's zero on all 6 external particles. Notice that $\mc{K}_{3,ij^\prime}$ is a $205\times 40$ matrix and we can join all six of the $\mc{K}_{m, ij^\prime}, {m=1\sim6}$, into a $205\times240$ matrix $\mc{K}$, which encodes the full constraints from Adler zero conditions. Entries of  $\mc{K}$ are usually highly dependent, and some of them are trivially zero. The number of independent constraints is obtained from the rank of $\mc{K}$ and the number of independent soft blocks $\mc{N}^{6,6}$ at $O(\phi^6 p^6)$ is given by 
\eq{\label{eq:adler_count}
	\mc{N}^{6,6} = \mathbb{d}_{\mc{M}}^{6,6} - {\rm rank}(\mc{K}) = 25.
}
Explicit forms of the soft blocks can be  obtained by solving the system of equations
\eq{\label{eq:adler_eqn}
	\sum_{i=1}^{205} \ c_i\, \mc{K}_{m,ij^\prime} = 0, \qquad m=1,\cdots,6\ ;\ \  j=1,\dots,40.
}
So that we obtain 25 independent solutions $c^{(1\sim25)}_i$ as the coordinates of the amplitudes $\mc{M}^{6,6}_{(1\sim25)} = \sum_ic^{(1\sim25)}_i\mc{M}^{\rm y}_i$ surviving the Adler zero constraints. 

The advantage of the general algorithm is that it directly provides a local amplitude satisfying Gram determinant redundancy. This is the algorithm implemented on \texttt{Mathematica} in the package ABC4EFT \cite{Li:2022tec}. External sources and fermions could be incorporated straightforwardly by introducing polarization tensors and additional spinor variables, as was done in Ref.~\cite{Sun:2022ssa}.
However, a drawback of the method is the outcome contains huge polynomials of the spinor brackets and is not in a form easy to use. The distinction between parity-even and parity-odd local amplitudes might not be discernible either. Moreover, for scalar interactions without external sources, we should be able to write the local amplitudes directly in terms of the kinematic invariants $s_{ij}$. In what follows, we will propose a second method to write the purely mesonic local amplitudes directly in terms of $s_{ij}$.

The spirit of the second algorithm is to construct an over-complete basis in monomials of $s_{ij}$, which satisfy the Adler's zero condition. In the example of the invariant operator $\phi^6D^6$, the monomials 
\eq{
s_{12}s_{34}s_{56} \qquad \text{and} \qquad   s_{12}\,\epsilon(3,4,5,6)\equiv s_{12}\, \epsilon_{\mu\nu\delta\gamma}p_3^\mu p_4^\nu p_5^\delta p_6^\gamma 
}
obviously satisfy the Adler's zero condition for every external leg at $O(p^6)$. Then we construct an over-complete basis for parity-even amplitudes, by permutating through  particle labels,
\eq{
\mc{M}^{\rm even}_i &\equiv \sigma_i\circ (s_{12}s_{34}s_{56}) = s_{\sigma_i(1)\sigma_i(2)}s_{\sigma_i(3)\sigma_i(4)}s_{\sigma_i(5)\sigma_i(6)}  \ ,\\
\mc{M}^{\rm odd}_i &\equiv \sigma_i\circ (s_{12}\,\epsilon(3,4,5,6)) = s_{\sigma_i(1)\sigma_i(2)}\,\epsilon(\sigma_i(3),\sigma_i(4),\sigma_i(5),\sigma_i(6)) \ ,
}
where $\sigma_i\in S_6$. Simple combinatorics show that there are a total of 15 $\mc{M}^{\rm even}$ and also 15 $\mc{M}^{\rm odd}$. These bases a priori do not take into account specific constraints in $D=4$ such as the Gram determinant. However, since amplitudes in the y-basis already included these constraints, we can simply project $\mc{M}^{\rm even}$ and $\mc{M}^{\rm odd}$ into the y-basis for 6-pt amplitudes, $\mc{M}^{\rm y}_i,\,i=1,\dots,205$, as $\mc{M}^{\rm even/odd}_i = \mc{K}^{\rm sy}_{ij} \mc{M}^{\rm y}_j$. The rest of the job is then to select the independent linear combinations of y-basis amplitudes from the projection. For example, we list the y-basis coordinates of three parity-even and one parity-odd amplitudes:
\eq{
	&s_{12}s_{34}s_{56} = -\mc{M}^{\rm y}_{23} +\mc{M}^{\rm y}_{38} -\mc{M}^{\rm y}_{39} -\mc{M}^{\rm y}_{45} -\mc{M}^{\rm y}_{55} +\mc{M}^{\rm y}_{56}, \\
	&s_{12}s_{35}s_{46} = \mc{M}^{\rm y}_{29} -\mc{M}^{\rm y}_{30} -\mc{M}^{\rm y}_{37} -\mc{M}^{\rm y}_{50} +\mc{M}^{\rm y}_{51} -\mc{M}^{\rm y}_{62}, \\
	&s_{13}s_{25}s_{46} = -\mc{M}^{\rm y}_{81} +\mc{M}^{\rm y}_{82} +\mc{M}^{\rm y}_{89} -\mc{M}^{\rm y}_{175} +\mc{M}^{\rm y}_{176} -\mc{M}^{\rm y}_{187}, \\
	&\epsilon(1,2,3,4)s_{56} = -\mc{M}^{\rm y}_{96} +\mc{M}^{\rm y}_{97} +\mc{M}^{\rm y}_{110} -\mc{M}^{\rm y}_{111}.
}
It turns out that, in this particular case, all 15 parity-even monomials constructed this way are independent, and 10 out of the 15 parity-odd amplitudes are independent. Their explicit forms are listed in Table.~\ref{tab:lorentz66_explicit} below. 
It is an amusing coincidence that for $\phi^6 D^6$, and more generally for $\phi^n D^n$, all monomials obtained through permutations in $\mc{M}^{\rm even}$ are independent. This is not the case in, for instance, $\phi^n D^{n+2}$.

\begin{table}[htbp]
\centering
\begin{align*}
\begin{array}{c|c}
\text{Parity even} & \text{Parity odd} \\
\hline
\mc{M}^{\rm even}_1=s_{14}s_{25}s_{36} & \mc{M}^{\rm odd}_{1}=s_{12}\epsilon(3,4,5,6) \\
\mc{M}^{\rm even}_2=s_{14}s_{26}s_{35} & \mc{M}^{\rm odd}_{2}=s_{13}\epsilon(2,4,5,6) \\
\mc{M}^{\rm even}_3=s_{15}s_{24}s_{36} & \mc{M}^{\rm odd}_{3}=s_{14}\epsilon(2,3,5,6) \\
\mc{M}^{\rm even}_4=s_{15}s_{26}s_{34} & \mc{M}^{\rm odd}_{4}=s_{15}\epsilon(2,3,4,6) \\
\mc{M}^{\rm even}_5=s_{16}s_{24}s_{35} & \mc{M}^{\rm odd}_{5}=s_{23}\epsilon(1,4,5,6) \\
\mc{M}^{\rm even}_6=s_{16}s_{25}s_{34} & \mc{M}^{\rm odd}_{6}=s_{24}\epsilon(1,3,5,6) \\
\mc{M}^{\rm even}_7=s_{13}s_{25}s_{46} & \mc{M}^{\rm odd}_{7}=s_{25}\epsilon(1,3,4,6) \\
\mc{M}^{\rm even}_8=s_{13}s_{26}s_{45} & \mc{M}^{\rm odd}_{8}=s_{34}\epsilon(1,2,5,6) \\
\mc{M}^{\rm even}_9=s_{15}s_{23}s_{46} & \mc{M}^{\rm odd}_{9}=s_{35}\epsilon(1,2,4,6) \\
\mc{M}^{\rm even}_{10}=s_{16}s_{23}s_{45}& \mc{M}^{\rm odd}_{10}=s_{45}\epsilon(1,2,3,6)\\
\mc{M}^{\rm even}_{11}=s_{13}s_{24}s_{56}\\
\mc{M}^{\rm even}_{12}=s_{14}s_{23}s_{56}\\
\mc{M}^{\rm even}_{13}=s_{12}s_{35}s_{46}\\
\mc{M}^{\rm even}_{14}=s_{12}s_{36}s_{45}\\
\mc{M}^{\rm even}_{15}=s_{12}s_{34}s_{56}\\
\end{array}
\end{align*}
\caption{Parity-even and parity-odd amplitude basis for soft blocks at $\phi^6D^6$.}
\label{tab:lorentz66_explicit}
\end{table}

This method is extremely easier than the general algorithm when the power of momenta is not much higher than the number of NGB. Otherwise, there may be too many possible kinematic factors to be inspected.

The coordinate of the kinematic factors under the y-basis is not only helpful for finding an independent set of monomials, but also provides a way to identify an arbitrary soft block as a combination of the soft basis amplitudes. Suppose we have an arbitrary kinematic factor $\mc{M} = \sum_i \mc{K}_i \mc{M}^{\rm y}_i$ with the unique coordinate $\mc{K}_i$ under the y-basis. We could try to solve the linear equations
\eq{\label{eq:soft_coord}
    \sum_{i} \mc{K}^{\rm soft}_i \mc{K}^{\rm sy}_{ij} = \mc{K}_j
}
to get the coordinate $\mc{K}^{\rm soft}_i$ of $\mc{M}$ under the soft basis $\mc{M}^{\rm even/odd}$. Of course, the equations have no solution if $\mc{M}$ does not actually satisfy the Adler zero condition.

\section{The Flavor Factor}
\label{sect:flavorfactor}

In the ChPT, the NGB transforms under the adjoint representation of the unbroken flavor group $H = SU(N_f)$. The operators must involve an invariant tensor with adjoint indices under $H$, called the flavor factor, which are usually written as traces over $T^a$, the $N_f\times N_f$ matrices satisfying the Lie algebra of $SU(N_f)$ group.
These traces are explicit $H$-invariant and easy to organize, but they are not always independent. In this section, we discuss the method of obtaining independent $H$-invariant tensors with rank given by the multiplicity of the NGB.

In general, the set of rank-$N$ independent invariant tensors can be obtained by taking the tensor product of $N$-copy of adjoint representations and focusing on the singlet representations in the decomposition. For example, for $SU(3)$ it is well-known that,
\eq{\label{eq:tpdadjoint}
    &\begin{array}{ccccccccccccc}
    \mathbf{8}&\times&\mathbf{8}&=& \mathbf{27} &+& \mathbf{10} &+& \mathbf{10} &+& \mathbf{8}\times2 &+& \mathbf{1}
    \end{array} \\
    &\begin{array}{ccccccccc}
    \mathbf{8}&\times&(\mathbf{8}&\times&\mathbf{8})&=& \mathbf{1}\times2 &+& \dots
    \end{array}
}
which implies there are 2 independent rank-3 invariant tensors in the adjoint of $SU(3)$, given by  precisely the structure constants $f^{abc}$ and the totally symmetric  $d^{abc}$. For general $SU(N_f)$ the rank-$N$ invariant tensors are the collection of Clebsch-Gordan coefficients for the tensor product decomposition of the representations of the fields \cite{Li:2020gnx,Li:2020xlh,Li:2022tec}, which can be computed repeatedly using the Littlewood-Richardson rule and  is implemented in the package  \textcolor{red}{\texttt{ABC4EFT}}. We list some results in the table~\ref{tab:nftensor} for reference.

\begin{table}[tbp]
	\centering
	\begin{tabular}{l|cccccc|r}
	    \hline
		Group						&	$SU(2)$	&	$SU(3)$	&	$SU(4)$	&	$SU(5)$	&	$SU(6)$	&	$SU(7)$	&	Trace \\
		\hline\hline
		$T^{a_1a_2a_3}$				&	\red{1}		&	2		&	2		&	2		&	2		&	2		&	2	\\
		$T^{a_1a_2a_3a_4}$			&	\red{3}		&	\red{8}		&	9		&	9		&	9		&	9		&	9	\\
		$T^{a_1a_2a_3a_4a_5}$		&	\red{6}		&	\red{32}		&	\red{43}		&	44		&	44		&	44		&	44	\\
		$T^{a_1a_2a_3a_4a_5a_6}$	&	\red{15}		&	\red{145}		&	\red{245}		&	\red{264}		&	265		&	265		&	265	\\
		\hline
	\end{tabular}
	\caption{The number of independent rank-$N$ invariant tensors for various groups $SU(N_f)$. In the last column, we also present the number of inequivalent trace structures, i.e., those not related by cyclic permutations. Cayley-Hamilton theorem reduces the number for $N_f<N$ as shown by the red numbers.}\label{tab:nftensor}
\end{table}

When the rank of the invariant tensor becomes larger than the number of flavors, $N>N_f$, one subtlety arises because linear relations among the tensors due to the Cayley-Hamilton theorem starts to appear and reduce the number of independent flavor structures. 
The Cayley-Hamilton theorem relates the $N_f^{\rm th}$ power of an $N_f\times N_f$ matrix to a polynomial of degree $N_f-1$ in itself. For instance, for $N_f=3$ we have
\eq{\label{eq:C-H_3}
	A^3 - \tr{A}A^2 + \frac12\left(\tr{A}^2 - \tr{A^2}\right)A - (\det A)\, I_{3\times3} = 0 \ ,
}
where $A$ is an arbitrary square matrix. 
If $\tr A = 0$, which is the case for the $SU(N_f)$ algebra, we can multiply $A$ to eq.~\eqref{eq:C-H_3} and take the trace to arrive at the relation:
\eq{\label{eq:tra4}
	\tr{A^4} = \frac12\tr{A^2}^2\ .
}
Now if we insert the linear combination  $A = \sum_a \alpha_a\mc{T}^a$, where  $\mc{T}^a$ are a basis for $SU(N_f)$ generators and $\alpha_a$'s are arbitrary constants, to eq.~\eqref{eq:tra4}, we arrive at many relations among the traces. For example, singling out the coefficient of $\alpha_a\alpha_b\alpha_c\alpha_d$, where $a, b, c,d$ are all distinct, we have
\eq{\label{eq:4pt_CH}
	\sum_{\sigma\in S_3}\tr{{T}^a{T}^{\sigma(b)}{T}^{\sigma(c)}{T}^{\sigma(d)}} = \tr{{T}^a{T}^b}\tr{{T}^c{T}^d} + \tr{{T}^a{T}^c}\tr{{T}^d{T}^b} + \tr{{T}^a{T}^d}\tr{{T}^b{T}^c},
}
which reduces the flavor structures starting at $N=4$ multiplicity for $N_f=3$.  

For $N_f\ge N$ when the Cayley-Hamilton does not apply, the trace structures (modulo a residual symmetry such as the cyclic permutations of the traces) are independent. Yet the subtlety is particularly pronounced in ChPT, where $N_f=2,3$ and $N$ could easily exceed $N_f$. In order to find the independent set of invariant tensors for small $N_f$,  we utilize the inner product between invariant tensors of the same rank introduced in Ref.~\cite{Li:2022tec},
\eq{\label{eq:tensorinner}
	(\mc{T}_i,\mc{T}_j) \equiv (\mc{T}_i^\dagger)_{a_1\dots a_N}(\mc{T}_j)^{a_1 \dots a_N} \ .
}
Notice the Hermitian conjugate inverts the traces, for example $ \tr{ABCD}^\dagger = \tr{DCBA} $, which renders the inner product  positive-definite $(\mc{T},\mc{T}) \ge 0$. From the inner products we define a metric tensor  if $\{\mc{T}_i\}$ forms a basis: $\bar{g}_{ij} \equiv (\mc{T}_i,\mc{T}_j)$. When $\{\mc{T}_i\}$ is not linearly independent, the metric is degenerate $\det \bar{g} = 0$. In this case a maximal independent subset $\mathfrak{T}$ can be obtained by demanding the corresponding minor metric $g_{ij} = \bar{g}_{ij},\ i,j\in\mathfrak{T}$, be non-singular $\det g \neq 0$. 

To see how this works explicitly, consider the following 9 different trace structures for 4-pt amplitudes,
\eq{\label{eq:4pt_tbasis}
	& \mc{T}_{i=1,\dots,6} = \tr{{T}^a{T}^{\sigma_i(b)}{T}^{\sigma_i(c)}{T}^{\sigma_i(d)}}, \quad \sigma_i \in S_3, \\
	& \mc{T}_7 = \tr{{T}^a{T}^b}\tr{{T}^c{T}^d} , \quad	\mc{T}_8 = \tr{{T}^a{T}^c}\tr{{T}^d{T}^b} , \quad	\mc{T}_9 = \tr{{T}^a{T}^d}\tr{{T}^b{T}^c} .
}
The inner product defined in eq.~\eqref{eq:tensorinner} can be computed by repeatedly using the orthogonality condition, 
\eq{\label{eq:Fierz}
{T}^a_{ij}{T}^a_{kl} = \delta_{il}\delta_{kj} - \frac{1}{N_f}\delta_{ij}\delta_{kl}.
}
For $N_f=3$, the resulting metric tensor  is
\eq{\label{eq:trace4_metric}
	\bar{g}_{ij}  = \frac{16}{3} \times \left(
	\begin{array}{ccccccccc}
		19 & -2 & -2 & -2 & -2 & 4 & 8 & -1 & 8 \\
		-2 & 19 & -2 & 4 & -2 & -2 & 8 & 8 & -1 \\
		-2 & -2 & 19 & -2 & 4 & -2 & -1 & 8 & 8 \\
		-2 & 4 & -2 & 19 & -2 & -2 & 8 & 8 & -1 \\
		-2 & -2 & 4 & -2 & 19 & -2 & -1 & 8 & 8 \\
		4 & -2 & -2 & -2 & -2 & 19 & 8 & -1 & 8 \\
		8 & 8 & -1 & 8 & -1 & 8 & 24 & 3 & 3 \\
		-1 & 8 & 8 & 8 & 8 & -1 & 3 & 24 & 3 \\
		8 & -1 & 8 & -1 & 8 & 8 & 3 & 3 & 24 \\
	\end{array}
	\right)_{9\times9}\, .
}
It turned out $\bar{g}_{ij}$ in eq.~\eqref{eq:trace4_metric} is a reduced rank, singular matrix, as expected; it has rank-8. We can find the redundancy relation by solving for the null space $\{\mathbf{n}^j\}$ of $\bar{g}$, which satisfies $\bar{g}_{ij} \mathbf{n}^j = 0$. For  eq.~\eqref{eq:trace4_metric} we find
\eq{
	\bar{g}_{ij} \mathbf{n}^j = 0 \qquad \Rightarrow \quad \mathbf{n}^iT_i = T_1+T_2+T_3+T_4+T_5+T_6-T_7-T_8-T_9=0,
}
which is precisely eq.~\eqref{eq:4pt_CH}, the redundancy relation obtained by using the Cayley-Hamilton theorem.
Consequently we can remove any one of the 9 redundant tensors to arrive at an independent tensor basis for $N_f=3$, with the metric $g$ being the corresponding minor of $\bar{g}$. Entries marked red in Table \ref{tab:nftensor} are the cases affected by the redundancy relations.

We can raise the index $i$ of an invariant tensor $\mc{T}_i$ by the metric tensor, 
\eq{
	\mc{T}^i = (g^{-1})^{ij}\mc{T}_j\ ,
}
such that for arbitrary tensor $\mc{T} = \sum_{i\in\mathfrak{T}} c_i \mc{T}_i$, we can extract the coordinates under the chosen basis $\{\mc{T}_{1,\dots,8}\}$ as
\eq{\label{eq:tensor_coord}
	c_i = (\mc{T}^i,\mc{T}) \,.
}
This technique turns out to be especially useful later. In general, the metric $\bar{g}_{ij}$  is a matrix dependent on $N_f$ due to the $N_f$ dependence in eq.~\eqref{eq:Fierz}. The rank of $\bar{g}_{ij}$  decreases as $N_f$ becomes smaller. For each $N_f$, we can choose an independent subset $\mathfrak{T}(N_f)$ as the basis of flavor group structures.

\section{Combining Kinematics with Flavors}
The full amplitude in ChPT is a product of kinematic invariants with the flavor factors. Due to the Bose-Einstein statistics, the full amplitude must be totally symmetric under simultaneous permutations of the kinematic indices and the flavor indices. 
In SMEFT this issue of spin-statistics was dealt with using the Clebsch-Gordan coefficients for  products of $S_n$ irreps, under which the  gauge group factors and the kinematic factors transform \cite{Li:2020gnx}. Another method uses  ``Young symmetrizers" for operator types with different repeated field patterns \cite{Li:2022tec,Sun:2022ssa}. In ChPT these methods are not computationally efficient because both the kinematic and flavor factors form  large linear spaces, as shown in Tables~\ref{tab:Gram} and \ref{tab:nftensor}. 
We are going to introduce yet another method more suitable for ChPT and demonstrate using the 6-pt $O(p^6)$ with $SU(3)$ flavor symmetry as an example.

Let us look at the parity-even soft blocks first, which contain 15 independent kinematic factors according to Table~\ref{tab:lorentz66_explicit}. The 6-pt trace structures for $SU(3)$ span a 145-dimensional space, as shown in Table~\ref{tab:nftensor}. Therefore the general space of possible parity-even amplitudes is $15\times 145 = 2175$-dimensional. The goal is to find the subspace containing totally symmetrized amplitudes so that Bose symmetry is respected. We can further divide the 145 flavor factors at 6-pt into 4 classes with different trace structures, and hence different permutation and cyclic symmetries. They are listed in Table~\ref{eq:trace6_summary}, where the last column presents the number of ``inequivalent traces'', meaning that they are not related by cyclic permutations within the traces. Note however the inequivalent traces may not be independent for the specific $SU(N_f)$ group; in fact there are 265 inequivalent traces at 6-pt, according to the last column of Table~\ref{tab:nftensor}, but only 145 of the 265 inequivalent traces are independent flavors factors for $N_f=3$.

\begin{table}[tbp] 
	\centering
	\begin{tabular}{|l|c|c|c|}
	    \hline
		\text{class}&	\text{typical form}																		& \text{short-hand notation}	&	\text{\#} \\
		\hline\hline
		(6)			&	$\tr{{T}^{a_1}{T}^{a_2}{T}^{a_3}{T}^{a_4}{T}^{a_5}{T}^{a_6}}$			& $\tr{123456}$ 	&	120	\\
		$(4|2)$		&	$\tr{{T}^{a_1}{T}^{a_2}{T}^{a_3}{T}^{a_4}}\tr{{T}^{a_5}{T}^{a_6}}$		& $\tr{1234|56}$	&	90	\\
		$(3|3)$		&	$\tr{{T}^{a_1}{T}^{a_2}{T}^{a_3}}\tr{{T}^{a_4}{T}^{a_5}{T}^{a_6}}$		& $\tr{123|456}$	&	40	\\
		$(2|2|2)$		&	$\tr{{T}^{a_1}{T}^{a_2}}\tr{{T}^{a_3}{T}^{a_4}}\tr{{T}^{a_5}{T}^{a_6}}$	& $\tr{12|34|56}$  &	15	\\
		\hline
	\end{tabular}
	\caption{Flavor factors for 6-pt amplitudes for $SU(N_f\ge 6)$. For $N_f<6$ additional redundancy relations appear, as explained in Section~\ref{sect:flavorfactor} and summarized in Table~\ref{tab:nftensor}.}
	\label{eq:trace6_summary}
\end{table}
 
The most direct way to generate full amplitudes satisfying the Bose symmetry is to introduce the  totally symmetric Young symmetrizer $\mc{Y}$, given explicitly by 
\eq{\label{eq:youngsym}
    \mc{Y} = \frac{1}{6!} \sum_{\sigma\in S_6} \sigma \,,
} 
where the permutations act on the amplitude as
\eq{
    \sigma \circ \left(\mc{T}^{a_1a_2\dots}\mc{M}(1,2,\dots)\right) = \mc{T}^{a_{\sigma(1)}a_{\sigma(2)},\dots}\mc{M}(\sigma(1),\sigma(2),\dots) \,.
}
Then applying the Young symmetrizer  $\mc{Y}$ to any product of kinematic and flavor factors would generate Bose-symmetric full amplitude. However, this is not the most efficient way to arrive at the desired outcome because the different product of kinematic and flavor factors could lead to the same full amplitude. A case in point is the two products, $\tr{123456} s_{12}$ and $\tr{123456} s_{23}$, which are equivalent under the action of $\mc{Y}$:
\eq{
	\mc{Y} \circ \left(\tr{123456} s_{12}\right) 
	&= \mc{Y} \circ \sigma(123456) \circ \left(\tr{123456} s_{12}\right) \\
		&= \mc{Y}  \circ \left(\tr{234561} s_{23}\right) \\
	&= \mc{Y} \circ \left(\tr{123456} s_{23}\right) \,,
}
where $(123456)$ represents the cyclic permutation of $i\to i+1 \mod 6$ and in the last equality we have used the cyclic invariance of the trace $\tr{234561}=\tr{123456}$.\footnote{From the definition of $\mc{Y}$ in eq.~\eqref{eq:youngsym} it should be clear that $\mc{Y}\circ \sigma =\mc{Y}$ for any $\sigma\in S_6$.} This redundancy results from the fact that the trace $\tr{234561}$ is invariant under the cyclic subgroup $H^{(6)}\simeq\mathbb{Z}_6$ of $S_6$ and that $s_{12}$ and $s_{23}$ are mapped into each other under the action of $H^{(6)}$,
\eq{
\sigma\circ s_{12} = s_{23} \ , \qquad\qquad \exists\;\sigma \in H^{(6)} \ .
}
In other words, as far as $\tr{123456}$ is concerned,  $s_{12}$ and $s_{23}$ belong to the same equivalent class under the action of $\mc{Y}$.

More generally, different trace structures are invariant under different subgroups of $S_6$. For a particular trace structure $T^{a_1a_2\dots}$ whose invariant group is $H$, two kinematic factors $\mc{M}_1$ and $\mc{M}_2$ would result in the same full amplitude under the action of $\mc{Y}$ if they are related to each other by the action of $H$.  These considerations motivate defining the  \emph{trace orbit} associated with  a particular subgroup $H_i$ of $S_6$:
\eq{
Orb_i(\mc{M}) \equiv H_i \circ\mc{M}\ ,
}
where $\mc{M}$ is a kinematic factor. 

For the parity-even soft blocks in Table~\ref{tab:lorentz66_explicit}, we find the following orbits for the  group $H^{(6)}$ which preserves the single trace $\tr{123456}$:
\eq{\label{eq:ex_orbit}
	& Orb_{H^{(6)}}(\mc{M}^{\rm even}_1) =  \mc{M}^{\rm even}_1 , \\
	& Orb_{H^{(6)}}(\mc{M}^{\rm even}_2) =  \{ \mc{M}^{\rm even}_2,\mc{M}^{\rm even}_{3},\mc{M}^{\rm even}_{7} \}, \\
	& Orb_{H^{(6)}}(\mc{M}^{\rm even}_4) =  \{ \mc{M}^{\rm even}_4,\mc{M}^{\rm even}_{5},\mc{M}^{\rm even}_{8},\mc{M}^{\rm even}_{9},\mc{M}^{\rm even}_{11},\mc{M}^{\rm even}_{13} \}, \\
	& Orb_{H^{(6)}}(\mc{M}^{\rm even}_6) =  \{ \mc{M}^{\rm even}_6,\mc{M}^{\rm even}_{12},\mc{M}^{\rm even}_{14} \}, \\
	& Orb_{H^{(6)}}(\mc{M}^{\rm even}_{10}) =  \{ \mc{M}^{\rm even}_{10},\mc{M}^{\rm even}_{15} \}.
}
There are five different trace orbits that sum up to the full space of 15 independent parity-even kinematic factors. Therefore, we expect five different full amplitudes satisfying Bose symmetry for the flavor factor $\tr{123456}$. When applying the Young symmetrizer $\mc{Y}$ in Eq.~\eqref{eq:youngsym}, it is sufficient to pick one element, or a linear combination, out of each trace orbit, for elements within the same orbit to give identical full amplitude upon the action of $\mc{Y}$. 

These five full amplitudes carrying the single trace structure are in one-to-one correspondence with the 6-pt soft blocks at $\mc{O}(p^6)$ computed in Ref.~\cite{Dai:2020cpk}. However, one crucial distinction is the soft blocks there are presented for the flavor-ordered, partial amplitudes,\footnote{Flavor-ordered partial amplitudes, similar to the color-ordered amplitudes in Yang-Mills theories, are amplitudes with the flavor factors stripped away \cite{Kampf:2012fn}.} which forms representations of the subgroup $H$ preserving a particular trace structure. In the case of $\tr{123456}$, Ref.~\cite{Dai:2020cpk} found five different soft blocks with the required cyclic invariance,  
\eq{\label{eq:symsoft}
    \mc{S}^{(6)}_1(1,2,3,4,5,6) &= \mc{M}_1^{\rm even}, \\
    \mc{S}^{(6)}_2(1,2,3,4,5,6) &= \frac{1}{3}\left( \mc{M}_2^{\rm even} + \mc{M}_3^{\rm even} + \mc{M}_7^{\rm even} \right) , \\
    \mc{S}^{(6)}_3(1,2,3,4,5,6) &= \frac{1}{6}\left( \mc{M}_4^{\rm even} + \mc{M}_5^{\rm even} + \mc{M}_8^{\rm even} + \mc{M}_9^{\rm even} + \mc{M}_{11}^{\rm even} + \mc{M}_{13}^{\rm even} \right) , \\
    \mc{S}^{(6)}_4(1,2,3,4,5,6) &= \frac{1}{3}\left( \mc{M}_6^{\rm even} + \mc{M}_{12}^{\rm even} + \mc{M}_{14}^{\rm even} \right) , \\
    \mc{S}^{(6)}_5(1,2,3,4,5,6) &= \frac{1}{2}\left( \mc{M}_{10}^{\rm even} + \mc{M}_{15}^{\rm even} \right) ,
}
which correspond to ``averaging'' over the element within each orbit in Eq.~\eqref{eq:ex_orbit}. In this work, we do not make this particular choice and instead will opt for choosing from each orbit a ``monomial'' of kinematic invariant in order to facilitate a direct mapping with operators in the Lagrangian formulation, as will be demonstrated later.

It is worth commenting on the observation that the union of the five orbits makes up the set of 15 parity even kinematic invariants for $\phi^6D^6$ operator. This appears to be a special case and not generic, which becomes evident when we consider the parity odd amplitudes in Table~\ref{tab:lorentz66_explicit}.

The trace orbit of parity odd amplitudes is a little trickier to compute because of the Levi-Civita tensor. For example, consider the action of  the cyclic permutation $h=\sigma(123456)$ on $\mc{M}^{\rm odd}_{4}$, 
\eq{\label{eq:modd4}
  h\circ\mc{M}^{\rm odd}_{4} = s_{26}\epsilon(3,4,5,1) 
}
which is outside of the basis of 10 independent parity-odd amplitudes chosen  in Table~\ref{tab:lorentz66_explicit}. To express Eq.~\eqref{eq:modd4} in terms of $\{\mc{M}^{\rm odd}_{1}, \cdots, \mc{M}^{\rm odd}_{10}\}$, the most systematic way is to project back to the y-basis and find the combination of $\mc{M}^{\rm odd}_i$ in terms of the coordinates,
\eq{
    s_{26}\epsilon(3,4,5,1) &= -\underbrace{\mc{M}^{\rm y}_{76} - \mc{M}^{\rm y}_{77} + \mc{M}^{\rm y}_{78} + \mc{M}^{\rm y}_{86} + \dots + \mc{M}^{\rm y}_{203}}_{\text{28 terms}} \nonumber \\
    &= \mc{M}^{\rm odd}_1 - \mc{M}^{\rm odd}_5 + \mc{M}^{\rm odd}_6 - \mc{M}^{\rm odd}_7\; ,
}
which can be done by the Mathematica code \hyperlink{https://abc4eft.hepforge.org/}{\texttt{ABC4EFT}} in milliseconds. In the end, we get 
\eq
{ \label{eq:traceodd1}
Orb(\mc{M}^{\rm odd}_{4}) = \left\{ \begin{array}{l} \mc{M}^{\rm odd}_{4}, \mc{M}^{\rm odd}_{1}-\mc{M}^{\rm odd}_{5}+\mc{M}^{\rm odd}_{6}-\mc{M}^{\rm odd}_{7} \\
 -\mc{M}^{\rm odd}_{2}, \mc{M}^{\rm odd}_{6}, -\mc{M}^{\rm odd}_{9}, \\
   - \mc{M}^{\rm odd}_{3}+\mc{M}^{\rm odd}_{6}-\mc{M}^{\rm odd}_{8}+\mc{M}^{\rm odd}_{10} 
     \end{array} \right\}.
    }
Proceed similarly to compute the orbit of $Orb(\mc{M}^{\rm odd}_{1})$,
\eq{\label{eq:traceodd2}
    & Orb(\mc{M}^{\rm odd}_{1}) = \left\{ \begin{array}{l} \mc{M}^{\rm odd}_{1}, -\mc{M}^{\rm odd}_{5}, \mc{M}^{\rm odd}_{8}, -\mc{M}^{\rm odd}_{10}, \\
    -\mc{M}^{\rm odd}_{4}+\mc{M}^{\rm odd}_{7}-\mc{M}^{\rm odd}_{9}+\mc{M}^{\rm odd}_{10}, \\
    -\mc{M}^{\rm odd}_{1}+\mc{M}^{\rm odd}_{2}-\mc{M}^{\rm odd}_{3}+\mc{M}^{\rm odd}_{4} \end{array} \right\}\ .
}
At this point we see the union of $Orb(\mc{M}^{\rm odd}_{1})$ and $Orb(\mc{M}^{\rm odd}_{4})$ is already larger than the set of 10 independent parity odd amplitudes in Table~\ref{tab:lorentz66_explicit}, unlike in the case of parity even amplitudes in Eq.~\eqref{eq:ex_orbit}. This suggests there are only 2 inequivalent parity odd full amplitudes, represented by $\mc{M}^{\rm odd}_{1,4}$,  under the multiplication of $\tr{123456}$ and total symmetrization, which correspond to two independent parity odd operators in the Lagrangian. 

\begin{table}[th]
    \centering{\small
    \begin{align*}
	\begin{array}{l|l}
		\hline\hline
		(6)\quad H[(123456)]      &                (4|2)\quad H[(1234),(56)]																			\\		
		\hline
		 \{ \mc{M}^{\rm even}_1 \}     &      \{\mc{M}^{\rm even}_1,\mc{M}^{\rm even}_2,\mc{M}^{\rm even}_4,\mc{M}^{\rm even}_6,\mc{M}^{\rm even}_{9},												\\		
		 \{ \mc{M}^{\rm even}_2,\mc{M}^{\rm even}_{3},\mc{M}^{\rm even}_{7} \}     &     	\qquad\quad \mc{M}^{\rm even}_{10},\mc{M}^{\rm even}_{13},\mc{M}^{\rm even}_{14}\}														\\		
		 \{ \mc{M}^{\rm even}_4,\mc{M}^{\rm even}_{5},\mc{M}^{\rm even}_{8},\mc{M}^{\rm even}_{9},\mc{M}^{\rm even}_{11},\mc{M}^{\rm even}_{13} \}        &      \{\mc{M}^{\rm even}_3,\mc{M}^{\rm even}_5,\mc{M}^{\rm even}_{7},\mc{M}^{\rm even}_{8}\}	\\		
		 \{ \mc{M}^{\rm even}_6,\mc{M}^{\rm even}_{12},\mc{M}^{\rm even}_{14} \}       &      \{\mc{M}^{\rm even}_{11}\}		\\
		 \{ \mc{M}^{\rm even}_{10},\mc{M}^{\rm even}_{15} \}     &   \{\mc{M}^{\rm even}_{12},\mc{M}^{\rm even}_{15}\} \\				
		\hline\hline
		(3|3)\quad H[(123),(456),(14)(25)(36)] 		&		(2|2|2)\quad H=[(12),(34),(56),(13)(24),(135)(246)] \\
		\hline
		\{\mc{M}^{\rm even}_1,\mc{M}^{\rm even}_4,\mc{M}^{\rm even}_5\}		& \{\mc{M}^{\rm even}_1,\mc{M}^{\rm even}_{2},\mc{M}^{\rm even}_{3},\mc{M}^{\rm even}_{5}, \\
		\{\mc{M}^{\rm even}_2,\mc{M}^{\rm even}_3,\mc{M}^{\rm even}_{6}\}		    & \qquad\quad \mc{M}^{\rm even}_{7},\mc{M}^{\rm even}_{8},\mc{M}^{\rm even}_{9},\mc{M}^{\rm even}_{10}\} \\
		\{\mc{M}^{\rm even}_7,\mc{M}^{\rm even}_8,\mc{M}^{\rm even}_9,\mc{M}^{\rm even}_{10},\mc{M}^{\rm even}_{11},		& \{\mc{M}^{\rm even}_4,\mc{M}^{\rm even}_6,\mc{M}^{\rm even}_{11},\mc{M}^{\rm even}_{12},\mc{M}^{\rm even}_{13},\mc{M}^{\rm even}_{14}\}  \\
		\qquad\quad \mc{M}^{\rm even}_{12},\mc{M}^{\rm even}_{13},\mc{M}^{\rm even}_{14},\mc{M}^{\rm even}_{15}\}      & \{\mc{M}^{\rm even}_{15}\} \\
		\hline
	\end{array}
	\end{align*}}
	\caption{Orbit spaces for 6-point $O(p^6)$ soft blocks, organized for each class of 6-point trace structures. The generators of the invariant subgroups $H$ are also provided in the square brackets.}\label{eq:orbit_space_2}
\end{table}

In the end, there are five different parity-even and two parity-odd distinct amplitudes with the single trace flavor structure at the $\phi^6D^6$ order. They are obtained by applying the Young symmetrizer $\mc{Y}$ to $\tr{123456}\mc{M}$, 
\eq{\label{eq:youngact}
\mc{Y}\circ \left(\tr{123456}\mc{\overline{M}}\right) \ , \quad \mc{\overline{M}} =\{\mc{M}^{\rm even}_1, \mc{M}^{\rm even}_2, \mc{M}^{\rm even}_4, \mc{M}^{\rm even}_6, \mc{M}^{\rm even}_{10}, \mc{M}^{\rm odd}_{1},\mc{M}^{\rm odd}_{4}\}\ ,
}
where $\mc{\overline{M}}$ consists of a monomial kinematic invariant from each of the orbits listed in Eqs.~\eqref{eq:ex_orbit}, \eqref{eq:traceodd1} and  \eqref{eq:traceodd2}. Notice that upon the action of the Young symmetrizer the particle labels $\{1, 2, \cdots\}$ become dummy and one could start with any one of 120 inequivalent single trace structures.\footnote{For example, one could have arrived at the same set of full amplitudes using $\mc{Y}\circ \left(\tr{213456}\mc{\overline{M}}\right)$ in Eq.~\eqref{eq:youngact}.}

The same procedure can be applied to the other trace classes, which have distinct invariant subgroup $H$, specified by the set of generators that include both the cycles of each trace and the permutations of traces with the same lengths. For example, the $(3|3)$ class of trace structures has a symmetry subgroup $H^{(3,3)}$ generated by the two cycles $(123)$ and $(456)$ and the swapping of the two traces $(14)(25)(36)$. With the generators, all the group elements can be obtained, which act on the amplitudes to find the orbits. In Table~\ref{eq:orbit_space_2} we list the parity-even orbits for all the 6-pt trace classes listed in Table~\ref{eq:trace6_summary}. 

Finally, we obtain a set of soft blocks that are independent without considering the finite $N_f$ effect, since we have only considered the symmetries of the traces. Therefore instead of ``independent'', we call them ``inequivalent'' soft blocks.

\subsection{Gram Determinant at $O(p^{10})$}

Before moving on, we want to emphasize that the Gram determinant has already been taken into account in the above analysis, because the spinor helicity formalism of the y-basis amplitudes is only valid at $D=4$.

At order $O(p^{10})$ constraints from the Gram determinant become active and reduce the number of independent amplitudes/operators at spacetime dimension $D=4$, which implies the following $n\times n$ matrix
\begin{equation}
\begin{pmatrix} 
s_{11} & s_{12} & \dots &s_{1n} \\
\vdots & \vdots & \ddots &\vdots \\
s_{1n} & s_{2n} &\dots  & s_{nn} 
\end{pmatrix},
\label{eq:gram}
\end{equation}
which has a rank of at most $D$, giving rise to the condition that any $(D+1)\times(D+1)$ sub-Gram matrix has zero determinant. The spinor helicity formalism of the y-basis amplitudes takes into account the Gram determinant.
To see that this is indeed the case, we compute some $O(p^{10})$ examples and compare them with those presented in Ref.~\cite{Dai:2020cpk} in general $D$.

The simplest example is for 6-pt at $O(p^{10})$, for which the number of independent soft blocks from  Ref.~\cite{Dai:2020cpk} is shown in the first row in Table~\ref{tab:sb_610}. In the second row, we present the numbers of independent orbit spaces with the help of the code \hyperlink{https://abc4eft.hepforge.org/}{\texttt{ABC4EFT}}. We get exactly one less amplitude/operator in $D=4$ than in the general $D$  for each trace structure. This can be easily understood because, at 6-pt, there is only one independent sub-Gram determinant after imposing momentum conservation, \eq{\label{eq:gram6}
   {\rm det}\ \left| \begin{array}{ccccc} 
                      s_{11} & s_{12} & s_{13} &  s_{14} & s_{15}  \\
                      s_{21} & s_{22} & s_{23} &  s_{24} & s_{25}  \\
                      s_{31} & s_{32} &  s_{33} &  s_{34} & s_{35}  \\
                      s_{41} & s_{42} & s_{43} &  s_{44} & s_{45}  \\
                      s_{51} & s_{52} & s_{53} &  s_{54} & s_{55} 
\end{array}    \right| = 0\,,
}
as those involving $p_6$ can always be converted to the above. The above equation also makes it clear that the determinant satisfies the Adler's zero condition for every external leg and is a soft block itself. Furthermore, the determinant is invariant under all permutations in $S_6$, which implies it forms a trace orbit on its own for every trace structure at the 6-pt level. As a result, setting the determinant to zero will remove exactly one trace orbit for every trace structure, as shown in Table~\ref{tab:sb_610}.

\begin{table}[t]
\centering
\begin{tabular}{|c|c|c|c|c|}
    \hline
    Trace Structure                         &   $(6)$    &   $(4|2)$   &   $(3|3)$   &   $(2|2|2)$  \\
    \hline
     General $D$ &   112  &   91  &   43  &   25  \\
    \hline
    $D=4$    &   111 &   90  &   42  &   24  \\
    \hline
\end{tabular}
\caption{Number of independent 6-pt amplitudes/operators at $O(p^{10})$ for each trace structure in general spacetime dimensions $D$ (given in \cite{Dai:2020cpk}) and specifically in $D=4$.}\label{tab:sb_610}
\end{table}

Going to 8-pt or higher, there are more multiple independent vanishing Gram determinants and, more importantly, they are not necessarily soft blocks anymore. The reduction in amplitude/operator basis due to constraints from the Gram determinant becomes less straightforward to count and nevertheless can be done using \hyperlink{https://abc4eft.hepforge.org/}{\texttt{ABC4EFT}}.
We show another example of 8- and 10-pt $O(p^{10})$ in Table~\ref{tab:sb_810} and~\ref{tab:sb_1010} with comparison to \cite{Dai:2020cpk}, where reduction of number of soft blocks vary for different trace structures.

\begin{table}[th]
\centering \footnotesize
\begin{tabular}{|c|c|c|c|c|c|c|c|}
    \hline
    &   $(8)$    &   $(6|2)$   &   $(5|3)$   &   $(4|4)$ &   $(4|2|2)$ &   $(3|3|2)$ &   $(2|2|2|2)$  \\
    \hline
    General $D$ &   435  &   320  &   226  &   129 &   149  &   117  & 26 \\
    \hline
    $D=4$    &   427 &   314  &   222  &  126  &  146 &    115 & 25  \\
    \hline
\end{tabular}
\caption{Number of independent 8-pt Soft Blocks at $O(p^{10})$ for general spacetime dimensions $D$ (given in \cite{Dai:2020cpk}), and specifically for $D=4$ using SSYT reductions.}\label{tab:sb_810}
\end{table}

\begin{table}[th]
\centering \footnotesize
\begin{tabular}{|c|c|c|c|c|c|c|}
    \hline
    &   $(10)$    &   $(8|2)$   &   $(7|3)$   &   $(6|4)$ &   $(5|5)$ &   $(6|2|2)$  \\
    \hline
    General $D$ &   105  &   74  &   45  &   50 &   29  &   37  \\
    \hline
    $D=4$    &   99 &   71  &   43  &   47  &   27 &    35  \\
    \hline\hline
    &   $(5|3|2)$    &   $(4|4|2)$    &   $(4|3|3)$    &   $(4|2|2|2)$    &   $(3|3|2|2)$    &   $(2|2|2|2|2)$  \\
    \hline
    General $D$ &   35  &   30  &   21  &   18  &   18  &   7  \\
    \hline
    $D=4$    &   35 &   28  &   20  &   17  &   17 &    6  \\
    \hline
\end{tabular}
\caption{Number of independent 10-pt Soft Blocks at $O(p^{10})$ for general spacetime dimensions $D$ (given in \cite{Dai:2020cpk}), and specifically for $D=4$ using SSYT reductions.}\label{tab:sb_1010}
\end{table}

\section{Example: 6-pt Soft Blocks at $O(p^6)$ with arbitrary $N_f$}

Having obtained the inequivalent soft blocks, we are finally ready to deal with the redundancy due to the finite $N_f$ effect. Currently, we do not have a better analytic method to treat it, such as from a thorough understanding of the Cayley-Hamilton theorem, which would be an interesting future study. Hence we do it by brute force, with the help of standard bases for both the kinematic factors $\mc{M}^{\rm even/odd}_i$ and flavor factors $\mc{T}_i$. Basically, we explicitly evaluate the Young symmetrizer in the inequivalent soft blocks, and find the coordinate of the resulting amplitude under the outer product of the standard bases $\mc{T}_i\mc{M}^{\rm even/odd}_j$. In particular, for the $N$-pt soft block $\mc{B} = \mc{Y}\circ (\mc{T} \mc{M})$ where $\mc{T}$ is some trace structure and $\mc{M}$ is some corresponding orbit space, we have
\eq{\label{eq:tot_sym}
    \mc{B} = \mc{Y}\circ (\mc{T}\mc{M}) &=
    \frac{1}{N!}\sum_{\sigma\in S_N} (\sigma\circ\mc{T})(\sigma\circ\mc{M}) \\
    &= \sum_{i=1}^{145}\sum_{j=1}^{\substack{15(\text{even})\\10(\text{odd})}} \underbrace{\left[\frac{1}{N!}\sum_{\sigma\in S_N}c_{i}(\sigma)\mc{K}^{\rm soft}_{j}(\sigma)\right]}_{\text{coordinate}} \times (\mc{T}_{i}\mc{M}^{\rm even/odd}_{j}),
}
where $c_i(\sigma) = (\mc{T}^i,\mc{\sigma}\circ \mc{T})$ as shown in eq.~\eqref{eq:tensor_coord} and $\mc{K}^{\rm soft}_j(\sigma)$ is obtained as in eq.~\eqref{eq:soft_coord} for $\sigma\circ \mc{M}$.
With the \texttt{Mathematica} package \href{https://abc4eft.hepforge.org/}{ABC4EFT} \cite{Li:2022tec}, it turns out to be computationally viable. Note that the $N_f$ dependence will come in the coefficients $c_i(\sigma)$ through the computation of the inner product \eqref{eq:Fierz}. Then it is straightforward to check the independence among the coordinates by linear algebra and select an independent subset from the inequivalent soft blocks. The selection has arbitrariness, of course, and here we choose to prioritize the appearance of the single trace structure in the final soft basis. The result for the 6-pt parity-even soft basis at order $O(p^6)$ is presented in Table~\ref{tab:evenamp66}.

\begin{table}[ht]
    \begin{equation*}\begin{array}{|c|c|c|c|c|l|l|}
    \hline
		\multicolumn{5}{|c|}{SU(N_f)} & \text{Operator Basis}  & \text{Amplitude Basis} \\
    \hline\hline
         & & & & \multirow{3}{*}{$SU(2)$} & \mc{O}_1=\langle u^{\mu}u^{\nu}u^{\rho}u_{\mu}u_{\nu}u_{\rho}\rangle & \mc{B}_1 = \mc{Y} \circ \tr{123456} s_{14}s_{25}s_{36}\\
         & & & & & \mc{O}_2=\langle u^{\mu}u^{\nu}u^{\rho}u_{\mu}u_{\rho}u_{\nu} \rangle & \mc{B}_2 = \mc{Y} \circ \tr{123456} s_{14}s_{26}s_{35}\\
         & & & & & \mc{O}_3=\langle u^{\mu}u^{\nu}u^{\rho}u_{\rho}u_{\mu}u_{\nu} \rangle  & \mc{B}_3 = \mc{Y} \circ \tr{123456} s_{15}s_{26}s_{34}\\
         \cline{5-7}
         & & & \multicolumn{2}{c|}{\multirow{5}*{$SU(3)$}} & \mc{O}_4=\langle u^{\mu}u^{\nu}u^{\rho}u_{\rho}u_{\nu}u_{\mu} \rangle & \mc{B}_4 = \mc{Y} \circ \tr{123456} s_{16}s_{25}s_{34} \\
         & & & \multicolumn{2}{c|}{~} & \mc{O}_5=\langle u^{\mu}u^{\nu}u_{\nu}u^{\rho}u_{\rho}u_{\mu} \rangle  & \mc{B}_5 = \mc{Y} \circ \tr{123456} s_{16}s_{23}s_{45}\\
         & & & \multicolumn{2}{c|}{~} & \mc{O}_6=\langle u^{\mu}u^{\nu}u^{\rho}u_{\mu}\rangle \langle u_{\nu}u_{\rho}\rangle  & \mc{B}_6 = \mc{Y} \circ \tr{1234|56} s_{14}s_{25}s_{36}\\
         & & & \multicolumn{2}{c|}{~} & \mc{O}_7=\langle u^{\mu}u^{\nu}u^{\rho}u_{\nu}\rangle \langle u_{\mu}u_{\rho}\rangle  & \mc{B}_7 = \mc{Y} \circ \tr{1234|56} s_{15}s_{24}s_{36}\\
         & & & \multicolumn{2}{c|}{~} & \mc{O}_8=\langle u^{\mu}u^{\nu}u_{\mu}u^{\nu}\rangle \langle u_{\rho}u_{\rho}\rangle  & \mc{B}_8 = \mc{Y} \circ \tr{1234|56} s_{13}s_{24}s_{56}\\
         \cline{4-7}
         & & \multicolumn{3}{c|}{\multirow{5}*{$SU(4)$}} & \mc{O}_9=\langle u^{\mu}u^{\nu}u_{\nu}u_{\mu}\rangle \langle u^{\rho}u_{\rho}\rangle   & \mc{B}_9 = \mc{Y} \circ \tr{1234|56} s_{14}s_{23}s_{56}   \\
		 & & \multicolumn{3}{c|}{~} & \mc{O}_{10}=\langle u^{\mu}u^{\nu}u^{\rho}\rangle \langle u_{\mu}u_{\nu}u_{\rho}\rangle  & \mc{B}_{10} = \mc{Y} \circ \tr{123|456} s_{14}s_{25}s_{36}    \\
		 & & \multicolumn{3}{c|}{~} & \mc{O}_{11}=\langle u^{\mu}u^{\nu}u^{\rho}\rangle \langle u_{\mu}u_{\rho}u_{\nu}\rangle  & \mc{B}_{11} = \mc{Y} \circ \tr{123|456} s_{14}s_{26}s_{35}    \\
		 & & \multicolumn{3}{c|}{~} & \mc{O}_{12}=\langle u^{\mu}u^{\nu}u_{\mu}\rangle \langle u^{\rho}u_{\nu}u_{\rho}\rangle  & \mc{B}_{12} = \mc{Y} \circ \tr{123|456} s_{13}s_{25}s_{46}   \\
		 & & \multicolumn{3}{c|}{~} & \mc{O}_{13}=\langle u^{\mu}u^{\rho}\rangle \langle u^{\nu}u_{\mu}\rangle \langle u_{\rho}u_{\nu}\rangle   & \mc{B}_{13} = \mc{Y} \circ \tr{12|34|56} s_{14}s_{25}s_{36}   \\
         \cline{3-7}  
         & \multicolumn{4}{c|}{SU(5)} & \mc{O}_{14}=\langle u^{\mu}u^{\nu}\rangle \langle u^{\rho}u_{\rho}\rangle \langle u_{\mu}u_{\nu}\rangle  & \mc{B}_{14} = \mc{Y} \circ \tr{12|34|56} s_{15}s_{26}s_{34}   \\
		 \cline{2-7}  
         \multicolumn{5}{|c|}{SU(N_f \ge 6)} & \mc{O}_{15}=\langle u^{\mu}u_{\mu}\rangle \langle u^{\nu}u_{\nu}\rangle \langle u^{\rho}u_{\rho}\rangle     & \mc{B}_{15} = \mc{Y} \circ \tr{12|34|56} s_{12}s_{34}s_{56}  \\
    \hline
    \end{array}\end{equation*}
    \caption{Independent 6-point soft basis for parity even at order $O(p^6)$. We show explicitly how the amplitude basis becomes linearly redundant when $N_f$ decreases. The corresponding operators are also given in terms of building block $u_\mu(x)$ by eq.~\eqref{eq:block_correspond}.}\label{tab:evenamp66}
\end{table}

In these Tables, the correspondence between the Lorentz structure of the operator and the particle labeling in the kinematic factor should be apparent and is the reason for the choice of a monomial in each trace orbit. 
The corresponding operators $\mc{O}_i$ are obtained by the translation eq.~\eqref{eq:block_correspond}.
In this context, the Young symmetrizer $\mc{Y}$ can be given a physical interpretation of performing the "Wick contraction" when computing the scattering amplitude of a particular operator insertion. 
We can even work out \emph{how} they become non-independent at lower $N_f$ by solving linear equations among the coordinates of the amplitudes. For example when $N_f=2$, we have
\eq{\label{eq:amp_relation_example}
    \begin{pmatrix} \mc{B}_4 \\ \mc{B}_5 \\ \mc{B}_6 \\ \mc{B}_7 \\ \mc{B}_8 \\ \mc{B}_9 \end{pmatrix}
    = \begin{pmatrix} 
    1 & 1 & -1 \\
    1 & 1 & -1 \\
    1 & 1 & 0 \\
    0 & 1 & 1 \\
    0 & 0 & 2 \\
    2 & 2 & -2
    \end{pmatrix}
    \times \begin{pmatrix} \mc{B}_1 \\ \mc{B}_2 \\ \mc{B}_3 \end{pmatrix} ,\quad
    \begin{pmatrix} \mc{B}_{10} \\ \mc{B}_{11} \\ \mc{B}_{12} \\ \mc{B}_{13} \\ \mc{B}_{14} \\ \mc{B}_{15} \end{pmatrix}
    = \begin{pmatrix} 
    0 & -1 & 1 \\
    0 & 1 & -1 \\
    0 & 0 & 0 \\
    1 & 2 & 1 \\
    2 & 2 & 0 \\
    4 & 4 & -4
    \end{pmatrix}
    \times \begin{pmatrix} \mc{B}_1 \\ \mc{B}_2 \\ \mc{B}_3 \end{pmatrix}
}
These can be derived only by introducing the additional trace redundancies for $SU(2)$, because otherwise, all of them should be independent. But many other redundancy relations must be involved as well, making it tedious to prove by hand. In our algorithm, they are obtained automatically by linear algebraic manipulation of the coordinates under the standard basis. Note that the relations eq.~\eqref{eq:amp_relation_example} cannot be directly turned into operator relations, because EOM may be involved in operator reductions, which do not show up on the amplitude side. In other words, the corresponding operator relations hold only up to EOM.

The same can be done for parity-odd amplitudes as shown in table~\ref{tab:oddamp66}, which was not given in \cite{Dai:2020cpk}.
\begin{table}[t]
    \begin{equation*}\begin{array}{|c|c|l|l|}
	\hline
		\multicolumn{2}{|c|}{SU(N_f)} & \text{Operator Basis} & \text{Amplitude Basis} \\
	\hline\hline
		 & \multirow{3}*{SU(3)} & \mc{O}_{16}=\langle u^{\mu}u_{\mu}u^{\nu}u^{\rho}u^{\sigma}u^{\xi}\rangle \epsilon_{\nu\rho\sigma\xi} & \mathcal{B}_{16}=\mathcal{Y}\circ \tr{123456}s_{12}\epsilon(3,4,5,6) \\
		 & & \mc{O}_{17}=\langle u^{\mu} u^{\nu}u_{\mu}u^{\rho}u^{\sigma}u^{\xi}\rangle \epsilon_{\nu\rho\sigma\xi} & \mathcal{B}_{17}=\mathcal{Y}\circ \tr{123456}s_{13}\epsilon(2,4,5,6) \\
		 & & \mc{O}_{18}=\langle u^{\mu} u^{\nu}u^{\rho}u^{\sigma}\rangle\langle u_{\mu}u^{\xi}\rangle \epsilon_{\nu\rho\sigma\xi} & \mathcal{B}_{18}=\mathcal{Y}\circ \tr{1234|56}s_{15}\epsilon(2,3,4,6) \\
		 \cline{2-4}
		 \multicolumn{2}{|c|}{SU(N_f \ge 4)} & \mc{O}_{19}=\langle u^{\mu}u_{\mu}u^{\nu}\rangle\langle u^{\rho}u^{\sigma}u^{\xi}\rangle \epsilon_{\nu\rho\sigma\xi} & \mathcal{B}_{19}=\mathcal{Y}\circ \tr{123|456}s_{12}\epsilon(3,4,5,6)\\
	\hline
	\end{array}\end{equation*}
	\caption{Independent 6-point soft basis for parity odd at order $O(p^6)$. How the amplitude basis become linearly redundant when $N_f$ decreases is explicitly shown. The corresponding operators are also given in terms of building block $u_\mu(x)$ by eq.~\eqref{eq:block_correspond}.}\label{tab:oddamp66}
\end{table}
The final results in table~\ref{tab:evenamp66} and \ref{tab:oddamp66} are summarized in Table~\ref{tab:66counting}.
\begin{table}[h]
    \centering
    $\begin{array}{|l|ccccc|}
		\hline
		\text{6-pt }O(p^6)	&	SU(2) & SU(3) & SU(4) & SU(5) & SU(6) \\
		\hline
		\text{P-even} & 3 & 8 & 13 & 14 & 15 \\
		\text{P-odd} & 0 & 3 & 4 & 4 & 4 \\
		\hline
	\end{array}$
    \caption{Number of independent 6-pt soft blocks at order $O(p^6)$, with both parities and different finite $N_f$ flavor symmetries.}\label{tab:66counting}
    $\begin{array}{|l|ccccc|}
		\hline
		\text{6-pt }O(p^8)	&	SU(2) & SU(3) & SU(4) & SU(5) & SU(6) \\
		\hline
		\text{P-even} & 9 & 40 & 68 & 74 & 76 \\
		\text{P-odd} & 2 & 20 & 33 & 35 & 35 \\
		\hline
	\end{array}$
    \caption{Number of independent 6-pt soft blocks at order $O(p^8)$, with both parities and different finite $N_f$ flavor symmetries.}\label{tab:68counting}
\end{table}
A more complicated example for a 6-point $O(p^8)$ soft base is provided in the appendix~\ref{app:basis68} with a summary in Table~\ref{tab:68counting}.
The $SU(2)$ and $SU(3)$ data in the summary agree with the Hilbert Series method \cite{Graf:2020yxt}.
We provide a sample code in the complementary material for the readers.

\section{Summary}

In this work, we provide a systematic method to construct the all-order ``soft basis'', an independent basis of soft blocks at spacetime dimension $D=4$ and finite flavor number $N_f$. It serves as the seed for the ``soft recursion'' of scattering amplitudes among Numbu-Goldstone bosons. The improvement compared to the literature \cite{Graf:2020yxt,Dai:2020cpk} is threefold: 
\begin{enumerate}
\item We eliminate the extra redundancy relations a) from the Gram determinant specifically introduced at $D=4$, appearing at $O(p^{10})$ by using the y-basis reduction technique \cite{Li:2020zfq,Li:2021tsq,Li:2022tec}; b) also from the Cayley-Hamilton relations for finite $N_f\;(<N)$ by numerically examining the inner product matrix among the trace structures. 

\item Parity odd soft blocks can be obtained on the same footing since the y-basis includes both parities.

\item We are able to obtain a nice monomial form of the final soft block, leading to the corresponding simple form of the operator basis for ChPT. We provide the explicit forms of operator basis at order $O(p^6)$ and $O(p^8)$ (in the appendix~\ref{app:basis68}) for $N=6$ NGB's and any flavor number $N_f$.

\end{enumerate}
It is usually not enough to get only the minimal operator/amplitude basis, because in actual applications we are often given an amplitude with different form from any one of the basis amplitudes, thus one still needs to do complicated operator conversions. For example, to do matching of effective operators, we compute the UV amplitude which should match to some combination of the IR operator basis. 
Our algorithm naturally provides a routine (such as eq.~\eqref{eq:tot_sym}) of decomposing any soft amplitude into a combination of our soft basis with a unique coordinate, so that the subtlety of operator conversions is systematically solved. 

\vspace{1cm}

\acknowledgements
This work is supported by the National Key Research and Development Program of China under Grant No. 2021YFC2203004. 
I. Low and M.-L. X. are supported by the U.S. Department of Energy under contracts No. DE-AC02-06CH11357 at Argonne.
J. S. is supported by the NSFC under Grants No. 12025507, No. 11690022, and No. 11947302, by the Strategic Priority Research Program and Key Research Program of Frontier Science of the Chinese Academy of Sciences (CAS) under Grants No. XDB21010200, No. XDB23010000, No. XDPB15, and No. ZDBS-LY-7003, and by the CAS Project for Young Scientists in Basic Research under Grant No. YSBR-006.

\appendix

\section{Group Theory and the Y-Basis}\label{app:math}

\subsection{Semi-Standard Young Tableau and Fock conditions}\label{app:fock}

In this section, we introduce some concepts from group theory \cite{ma2007group}. A permutation $\sigma = (a_1\; a_2\;\cdots\;a_r)$ is defined to permute the tensor indices, for example,
\begin{align}
    (1\;2\;3) \circ T_{ijkl}=T_{kijl},
\end{align}
where $\sigma$ moves the indices from the $(a_i)$th to the $(a_{i+1\,{\rm mod}\,r})$th position. Such permutations form the symmetric group $S_r$, where the group product is given by the composition of permutations. 

In the group theory, we use the Young Diagram (YD), Young Tableau and Young operator to obtain the independent basis of tensors with particular symmetry. A Young Diagram $Y^{[\eta]}$ is made up of some cells, in terms of partition $[\eta]=[\eta_1,\eta_2,\cdots,\eta_m]$, where $\eta_1\ge\eta_2\ge\cdots\ge\eta_m$, $\sum_i^m\eta_i=r$. $\eta_i$ denotes the number of cells in the $i$th row, such as $Y^{[3,2]}={\tiny\yng(3,2)}$. An $S_r$ Young Tableau is obtained by filling in the Young Diagram with numbers 1 to $r$. Each row of such a Young Tableau represents a totally symmetric permutation, such as
\begin{align}\label{eq:EvenPerm}
    \young(123)\Rightarrow 1+(1\;2)+(1\;3)+(2\;3)+(1\;2\;3)+(3\;2\;1).
\end{align}
Each column of such a young Tableau represents an anti-symmetric permutation, such as
\begin{align}\label{eq:OddPerm}
    \young(1,2,3)\Rightarrow 1-(1\;2)-(1\;3)-(2\;3)+(1\;2\;3)+(3\;2\;1).
\end{align}
The Young Operator $\mc{Y}$ corresponding to $S_r$ Young Tableau is defined as the production of all kinds of eqs.~(\ref{eq:EvenPerm}-\ref{eq:OddPerm}). For example,
\begin{align}
\begin{split}\label{eq:ExampleYO}
    \mc{Y}\left[\young(123,45)\right]=&\frac{1}{\prod_{ij} h_{ij}}\big\{1+(1\;2)+(1\;3)+(2\;3)+(1\;2\;3)+(3\;2\;1)\big\}\\
    &\times\big\{1+(4\;5)\big\}\big\{1-(1\;4)\big\}\big\{1-(2\;5)\big\}\;,
\end{split}
\end{align}
where $h_{ij}$ is called the hook number, defined for each cell. The hook number $h_{ij}$ for the cell in the $i$th row and $j$th column is equal to the number of cells below plus the number of cells on the right plus 1. For example, the hook numbers in eq.~(\ref{eq:ExampleYO}) is below,
\begin{align}
    \begin{array}{c}
         \text{hook numbers in}  \\
         \text{Young Tableau} 
    \end{array}:\qquad \young(431,21)\;,\qquad \prod_{ij}h_{ij}=4\times3\times2=24\;.
\end{align}
Also the dimension of a $[\eta]$ representation space is obtained from hook numbers, $d_{[\eta]}=r!/\prod h_{ij}$.

Then we can define the $SU(N)$ Young Tableau (YT) by applying the $S_r$ Young Operator to the rank-$r$ $SU(N)$ tensor, and fill the Young Tableau by the corresponding indices. For example, 
\begin{equation}\label{eq:SUN_YT}
\begin{array}{ccll}
    SU(N)\text{ YT} & & S_r\text{ YT} &  SU(N)\text{ tensor} \\
    \young(ik,jl)&=&\mc{Y}\left[\young(13,24)\right]&\circ\; T_{ijkl}\;.
\end{array}
\end{equation}
where $i,j,k,l$ take values from 1 to $N$. The resulting tensor is also called an irreducible tensor under the irreducible representation given by the YD, in this case $\tiny \yng(2,2)$.
For this particular YD, the $r^N$ YTs are not independent due to the Fock conditions of the Young operator.
Suppose the $j$th column and $j'$th column of Young Diagram $\mc{Y}$ has $\tau$ and $\tau'$ cells, respectively, and $\tau\ge \tau'$. The numbers that fill these two columns in a Young tableau are $a_{i}$ and $b_{i'}$. Then the Fock conditions of Young Operator are
\begin{align}
    \mc{Y}\left(1-\sum^{\tau}_{i=1}(a_{i}\; b_{i'}) \right)=0,
\end{align}
where $(a_{i}\; b_{i'})$ denotes a permutation between $a_{i}$ and $b_{i'}$. The conditions applied to the Young tableau amplitudes will have the following equation.
\begin{align}\label{eq:fock}
    \young({}{{a_1}}{{}}{{}},{}{{a_2}}{{b_2}}{{}},{}{{a_3}})= \young({}{{b_2}}{{}}{{}},{}{{a_2}}{{a_1}}{{}},{}{{a_3}})+ \young({}{{a_1}}{{}}{{}},{}{{b_2}}{{a_2}}{{}},{}{{a_3}})+ \young({}{{a_1}}{{}}{{}},{}{{a_2}}{{a_3}}{{}},{}{{b_2}})\;.
\end{align}
In the end, the independent YTs can be obtained by the Semi-Standard Young Tableau (SSYT) characterized by the following features:
\begin{enumerate}
    \item Along each row, the labels are non-decreasing from left to right.
    \item Along each column, the labels are increasing from top to bottom.
\end{enumerate}
For example, for $SU(3)$ YT in the irreducible representation $\tiny \yng(2,1)$, there are 8 SSYTs as follows:
\eq{
    \young(11,2),\ \young(11,3),\ \young(12,2),\ \young(12,3),\ 
    \young(13,2),\ \young(13,3),\ \young(22,3),\ \young(23,3).
}
The dimension of $SU(N)$ representation space can be obtained directly by the \emph{Hook content formula}
\begin{align}\label{eq:HCF}
    \mc{S}([\eta],N)=\prod_{ij}\frac{N-i+j}{h_{ij}}.
\end{align}
The non-SSYT can be related to the SSYT by the total anti-symmetry of each column and the Fock condition such as
\eq{
    \young(31,2) \ =\ \young(13,2) + \young(32,1) \ =\ \young(13,2) - \young(12,3)\;.
}

\subsection{Young Tensor Method and the Amplitude Y-Basis}\label{app:y-basis}

We briefly introduce the idea of Young tensor method to construct the amplitude y-basis, initially proposed by Henning and Melia \cite{Henning:2019enq,Henning:2019mcv} and then developed into the current form by Li {\it etc.} \cite{Li:2020xlh,Li:2020gnx,Li:2022tec}. The idea is to define an $SU(N)$ group acting on the helicity-spinor variables as
\eq{
    \lambda_i \to \mc{U}_i^{\,j}\lambda_j\ ,\quad \tilde\lambda^i \to \tilde\lambda^j\mc{U}^{\dagger i}_j\ ,
    \qquad \mc{U}\in SU(N)\,.
}
As a result, the $N$-point on-shell amplitude transforms as an $SU(N)$ tensor. In particular,
\eq{
    \vev{ij} \in \yng(1,1)\ ,\quad [ij] \in  \left.\begin{array}{c}
\yng(1,1)\\
\vdots\\
\yng(1,1) 
\end{array}\right\} \rotatebox[]{-90}{\text{$N-2$}}.
}
The redundancy relations among the amplitudes, such as the momentum conservation and the Schouten identities, indicate symmetries among the tensor components, picking out particular representation spaces that survive the constraints. They are defined as primary representation, as defined in the main text. Therefore, the correspondence between the YT and the amplitudes is given by the definition of the $SU(N)$ YT in eq.~\eqref{eq:SUN_YT}. For example, the $N=4$ and $(\tilde{n},n)=(0,2)$ tensor $T_{ijkl}=\vev{ij}\vev{kl}$, 
\begin{align}\begin{split}
    \young(ik,jl)\simeq&\frac{1}{3\times2\times2}\big\{1+(1\;3)\big\}\big\{1+(2\;4)\big\}\big\{1-(1\;2)\big\}\big\{1-(3\;4)\big\}T_{ijkl}\\
    =&\frac{1}{12}\big\{1+(1\;3)+(2\;4)+(1\;3)(2\;4)\big\}( T_{ijkl}-T_{jikl}-T_{ijlk}+T_{jilk} )\\
    =&\frac{1}{12}\big(8\vev{ij}\vev{kl}+4\vev{ik}\vev{jl}-4\vev{il}\vev{jk} \big)\\
    =&\vev{ij}\vev{kl}\;.
\end{split}\end{align}
The same is true for square bracket, such as in the case $(\tilde{n},n)=(2,0)$
\begin{align}
    \young({{\color{blue}i}}{{\color{blue}k}},{{\color{blue}j}}{{\color{blue}l}})\simeq\mc{E}^{iji'j'}[i'j']\mc{E}^{klk'l'}[k'l']\,.
\end{align}
The correspondence is an isomorphism since the Fock conditions among the YT exactly correspond to the momentum conservation and the Schouten identity. 
More precisely, when the $a_{i}$ and $b_{i'}$ in eq.~\eqref{eq:fock} come from columns representing square bracket and angular bracket, respectively, the Fock conditions are equivalent to momentum conservation,
\begin{align}
    \young({{\color{blue}2}}1,{{\color{blue}3}}4)=\young({{\color{blue}1}}2,{{\color{blue}3}}4)+\young({{\color{blue}2}}3,{{\color{blue}1}}4) \quad\Longleftrightarrow\quad 
    [14]\vev{14}=-[24]\vev{24}-[34]\vev{34}\;.
\end{align}
When $a_{i}$ and $b_{i'}$ come from columns both representing square brackets or angular brackets, the Fock conditions are equivalent to Schouten identities. For example,
\begin{align}
    \young({{2}}1,{{3}}4)=\young({{1}}2,{{3}}4)+\young({{2}}3,{{1}}4) \quad\Longleftrightarrow\quad 
    \vev{23}\vev{14}=\vev{13}\vev{24}-\vev{12}\vev{34}\;.
\end{align}
Therefore the independent amplitudes are naturally given by the SSYT, namely the y-basis.

Meanwhile, we are able to convert any amplitudes the correspond to non-SSYT, towards SSYT amplitudes(y-basis) using momentum conservation and Schouten identity. It requires a certain order of the constituting fields, and different orders would lead to different y-basis. We usually adopt the helicity-non-decreasing order proposed in \cite{Henning:2019enq}, but it is not mandatory. 
 
\subsection{Y-Basis Reduction Rules and the Vanishment of Gram Determinants}
\label{append:gram} 

Applying the following rules to any amplitude in terms of spinor helicity variables could convert it into a combination of the y-basis defined above:
    \begin{itemize}
        \item Replace all the momenta of first particle by momentum conservation
            \begin{align}
                \langle i1\rangle[1j]=-\sum^N_{k=2}\langle ik\rangle[kj].\label{eq:rulep1}
            \end{align}
    
        \item Replace all the momenta of particle 2 or 3 in the following cases such that no lower label momenta would be generated,
            \eq{\label{eq:rulep3}
                & [1|p_2|i\rangle=-\sum_{k=3}^N[1|k|i\rangle,\quad \langle 1|p_2|i]=-\sum_{k=3}^N\langle 1|p_k|i],   \\
                & [1|p_3|2\rangle=-\sum_{k=4}^N[1|k|2\rangle,\quad \langle 1|p_3|2]=-\sum_{k=4}^N\langle 1|p_k|2], \\
                & p_2\cdot p_3=\sum_{\substack{i,j\neq 1\\ \{i,j\}\neq\{2,3\} }} -p_i\cdot p_j .
            }
            
        \item Second, when two same-type brackets contain 4 different particles with the order $i<j<k<l$, we use the Schouten Identity to apply the following replacement
    \begin{align}
        & \langle il\rangle\langle jk\rangle = \langle ik\rangle\langle jl\rangle - \langle ij\rangle\langle kl\rangle, \label{eq:ASch}\\
        & [il][jk] = [ik][jl] - [ij][kl].\label{eq:SSch}
    \end{align}
    \end{itemize}
Since the spinor formulation of amplitudes implies the $D=4$ nature, it can be verified that applying the above rules, in particular the Schouten identity, to the Gram determinant in eqs.~(\ref{eq:D4_even}-\ref{eq:D4_odd}) renders zero. Therefore, this redundancy does not appear in our y-basis formulation.

\section{Order $O(p^8)$ Amplitude/Operator Basis for ChPT}
\label{app:basis68}

\begin{table}[htbp]
	\centering 
\begin{align*} \footnotesize
    \begin{array}{|c|c|c|c|c|l|l|}
        \hline
		\multicolumn{5}{|c|}{SU(N_f)} & \text{Operator Basis}  & \text{Amplitude Basis} \\
        \hline\hline
         & & & & \multirow{9}*{$SU(2)$} & \mc{O}_1=\langle \nabla^{\mu}\nabla^{\nu}u^{\rho}u^{\sigma}u_{\mu}u_{\nu}u_{\rho}u_{\sigma}\rangle & \mc{B}_1 = \mc{Y} \circ \tr{123456} s_{13}s_{14}s_{15}s_{26}\\
         & & & & & \mc{O}_2=\langle \nabla^{\mu}\nabla^{\nu}u^{\rho}u^{\sigma}u_{\mu}u_{\nu}u_{\sigma}u_{\rho}\rangle & \mc{B}_2 = \mc{Y} \circ \tr{123456} s_{13}s_{14}s_{16}s_{25}\\
         & & & & & \mc{O}_3=\langle \nabla^{\mu}\nabla^{\nu}u^{\rho}u_{\mu}u^{\sigma}u_{\nu}u_{\sigma}u_{\rho} \rangle  & \mc{B}_3 = \mc{Y} \circ \tr{123456} s_{12}s_{14}s_{16}s_{35}\\
		 & & & & & \mc{O}_4=\langle \nabla^{\mu}u^{\nu}\nabla^{\rho}u^{\sigma}u_{\mu}u_{\nu}u_{\rho}u_{\sigma} \rangle  & \mc{B}_4 = \mc{Y} \circ \tr{123456} s_{13}s_{14}s_{25}s_{26}\\
		 & & & & & \mc{O}_5=\langle \nabla^{\mu}u^{\nu}\nabla^{\rho}u^{\sigma}u_{\mu}u_{\rho}u_{\nu}u_{\sigma} \rangle  & \mc{B}_5 = \mc{Y} \circ \tr{123456} s_{13}s_{15}s_{24}s_{26}\\
		 & & & & & \mc{O}_6=\langle \nabla^{\mu}u^{\nu}\nabla^{\rho}u^{\sigma}u_{\rho}u_{\mu}u_{\sigma}u_{\nu} \rangle  & \mc{B}_6 = \mc{Y} \circ \tr{123456} s_{14}s_{16}s_{23}s_{25}\\
		 & & & & & \mc{O}_7=\langle \nabla^{\mu}u^{\nu}\nabla_{\mu}u^{\rho}u^{\sigma}u_{\nu}u_{\rho}u_{\sigma} \rangle  & \mc{B}_7 = \mc{Y} \circ \tr{123456} s_{12}s_{14}s_{25}s_{36}\\
		 & & & & & \mc{O}_8=\langle \nabla^{\mu}u^{\nu}\nabla_{\mu}u^{\rho}u^{\sigma}u_{\nu}u_{\sigma}u_{\rho} \rangle  & \mc{B}_8 = \mc{Y} \circ \tr{123456} s_{12}s_{14}s_{26}s_{35}\\
		 & & & & & \mc{O}_9=\langle \nabla^{\mu}u^{\nu}\nabla_{\mu}u^{\rho}u^{\sigma}u_{\rho}u_{\nu}u_{\sigma} \rangle  & \mc{B}_9 = \mc{Y} \circ \tr{123456} s_{12}s_{15}s_{24}s_{36}\\
         \cline{5-7}
         & & & \multicolumn{2}{c|}{\multirow{31}*{$SU(3)$}} & \mc{O}_{10}=\langle \nabla^{\mu}\nabla^{\nu}u^{\rho}u^{\sigma}u_{\sigma}u_{\mu}u_{\nu}u_{\rho} \rangle & \mc{B}_{10} = \mc{Y} \circ \tr{123456} s_{14}s_{15}s_{16}s_{23} \\
         & & & \multicolumn{2}{c|}{~} & \mc{O}_{11}=\langle \nabla^{\mu}\nabla^{\nu}u^{\rho}u^{\sigma}u_{\mu}u_{\sigma}u_{\nu}u_{\rho} \rangle  & \mc{B}_{11} = \mc{Y} \circ \tr{123456} s_{13}s_{15}s_{16}s_{24}\\
         & & & \multicolumn{2}{c|}{~} & \mc{O}_{12}=\langle \nabla^{\mu}u^{\nu}\nabla^{\rho}u^{\sigma}u_{\rho}u_{\sigma}u_{\mu}u_{\nu} \rangle  & \mc{B}_{12} = \mc{Y} \circ \tr{123456} s_{15}s_{16}s_{23}s_{24}\\
         & & & \multicolumn{2}{c|}{~} & \mc{O}_{13}=\langle \nabla^{\mu}u^{\nu}\nabla^{\rho}u^{\sigma}u_{\mu}u_{\rho}u_{\sigma}u_{\nu} \rangle  & \mc{B}_{13} = \mc{Y} \circ \tr{123456} s_{13}s_{16}s_{24}s_{25}\\
         & & & \multicolumn{2}{c|}{~} & \mc{O}_{14}=\langle \nabla^{\mu}u^{\nu}\nabla^{\rho}u^{\sigma}u_{\rho}u_{\mu}u_{\nu}u_{\sigma} \rangle  & \mc{B}_{14} = \mc{Y} \circ \tr{123456} s_{14}s_{15}s_{23}s_{26}\\
		 & & & \multicolumn{2}{c|}{~} & \mc{O}_{15}=\langle \nabla^{\mu}\nabla^{\nu}u^{\rho}u_{\mu}u^{\sigma}u_{\sigma}u_{\nu}u_{\rho} \rangle  & \mc{B}_{15} = \mc{Y} \circ \tr{123456} s_{12}s_{15}s_{16}s_{34}\\
		 & & & \multicolumn{2}{c|}{~} & \mc{O}_{16}=\langle \nabla^{\mu}u^{\nu}u^{\rho}u^{\sigma}\nabla_{\rho}u_{\sigma}u_{\mu}u_{\nu} \rangle  & \mc{B}_{16} = \mc{Y} \circ \tr{123456} s_{15}s_{16}s_{24}s_{34}\\
		 & & & \multicolumn{2}{c|}{~} & \mc{O}_{17}=\langle \nabla^{\mu}u^{\nu}\nabla_{\mu}u^{\rho}u^{\sigma}u_{\sigma}u_{\rho}u_{\nu} \rangle  & \mc{B}_{17} = \mc{Y} \circ \tr{123456} s_{12}s_{16}s_{25}s_{34}\\
		 & & & \multicolumn{2}{c|}{~} & \mc{O}_{18}=\langle \nabla^{\mu}u^{\nu}\nabla_{\mu}u^{\rho}u^{\sigma}u_{\sigma}u_{\nu}u_{\rho} \rangle  & \mc{B}_{18} = \mc{Y} \circ \tr{123456} s_{12}s_{15}s_{26}s_{34}\\
		 & & & \multicolumn{2}{c|}{~} & \mc{O}_{19}=\langle \nabla^{\mu}u^{\nu}u^{\rho}\nabla_{\mu}u^{\sigma}u_{\sigma}u_{\nu}u_{\rho} \rangle  & \mc{B}_{19} = \mc{Y} \circ \tr{123456} s_{13}s_{15}s_{26}s_{34}\\
		 & & & \multicolumn{2}{c|}{~} & \mc{O}_{20}=\langle \nabla^{\mu}u^{\nu}u^{\rho}u^{\sigma}u_{\sigma}\nabla_{\mu}u_{\nu}u_{\rho} \rangle  & \mc{B}_{20} = \mc{Y} \circ \tr{123456} s_{15}^2s_{26}s_{34}\\
		 & & & \multicolumn{2}{c|}{~} & \mc{O}_{21}=\langle \nabla^{\mu}u^{\nu}\nabla_{\mu}u^{\rho}u^{\sigma}u_{\rho}u_{\sigma}u_{\nu} \rangle  & \mc{B}_{21} = \mc{Y} \circ \tr{123456} s_{12}s_{16}s_{24}s_{35}\\
		 & & & \multicolumn{2}{c|}{~} & \mc{O}_{22}=\langle \nabla^{\mu}u^{\nu}u^{\rho}\nabla_{\mu}u^{\sigma}u_{\nu}u_{\sigma}u_{\rho} \rangle  & \mc{B}_{22} = \mc{Y} \circ \tr{123456} s_{13}s_{14}s_{26}s_{35}\\
		 & & & \multicolumn{2}{c|}{~} & \mc{O}_{23}=\langle \nabla^{\mu}u^{\nu}u^{\rho}u^{\sigma}\nabla_{\mu}u_{\nu}u_{\sigma}u_{\rho} \rangle  & \mc{B}_{23} = \mc{Y} \circ \tr{123456} s_{14}^2s_{26}s_{35}\\
		 & & & \multicolumn{2}{c|}{~} & \mc{O}_{24}=\langle \nabla^{\mu}\nabla^{\nu}u^{\rho}u_{\mu}u^{\sigma}u_{\nu}u_{\rho}u_{\sigma} \rangle  & \mc{B}_{24} = \mc{Y} \circ \tr{123456} s_{12}s_{14}s_{15}s_{36}\\
		 & & & \multicolumn{2}{c|}{~} & \mc{O}_{25}=\langle \nabla^{\mu}u^{\nu}u^{\rho}\nabla_{\mu}u^{\sigma}u_{\rho}u_{\nu}u_{\sigma} \rangle  & \mc{B}_{25} = \mc{Y} \circ \tr{123456} s_{13}s_{15}s_{24}s_{36}\\
		 & & & \multicolumn{2}{c|}{~} & \mc{O}_{26}=\langle \nabla^{\mu}u^{\nu}u^{\rho}\nabla_{\mu}u^{\sigma}u_{\nu}u_{\rho}u_{\sigma} \rangle  & \mc{B}_{26} = \mc{Y} \circ \tr{123456} s_{13}s_{14}s_{25}s_{36}\\
		 & & & \multicolumn{2}{c|}{~} & \mc{O}_{27}=\langle \nabla^{\mu}u^{\nu}\nabla_{\mu}u^{\rho}u_{\rho}u^{\sigma}u_{\sigma}u_{\nu}\rangle  & \mc{B}_{27} = \mc{Y} \circ \tr{123456} s_{12}s_{16}s_{23}s_{45}\\
		 & & & \multicolumn{2}{c|}{~} & \mc{O}_{28}=\langle \nabla^{\mu}\nabla^{\nu}u^{\rho}u_{\mu}u_{\nu}u^{\sigma}u_{\rho}u_{\sigma} \rangle  & \mc{B}_{28} = \mc{Y} \circ \tr{123456} s_{12}s_{13}s_{15}s_{46}\\
		 & & & \multicolumn{2}{c|}{~} & \mc{O}_{29}=\langle \nabla^{\mu}\nabla^{\nu}u^{\rho}u_{\mu}u_{\nu}u_{\rho}u^{\sigma}u_{\sigma} \rangle  & \mc{B}_{29} = \mc{Y} \circ \tr{123456} s_{12}s_{13}s_{14}s_{56}\\
		 & & & \multicolumn{2}{c|}{~} & \mc{O}_{30}=\langle \nabla^{\mu}\nabla^{\nu}u^{\rho}u^{\sigma}u_{\mu}u_{\sigma}\rangle \langle u_{\nu}u_{\rho}\rangle  & \mc{B}_{30} = \mc{Y} \circ \tr{1234|56} s_{13}s_{15}s_{16}s_{24}\\
		 & & & \multicolumn{2}{c|}{~} & \mc{O}_{31}=\langle \nabla^{\mu}u^{\nu}\nabla^{\rho}u^{\sigma}u_{\rho}u_{\sigma}\rangle \langle u_{\mu}u_{\nu}\rangle  & \mc{B}_{31} = \mc{Y} \circ \tr{1234|56} s_{15}s_{16}s_{23}s_{24}\\
		 & & & \multicolumn{2}{c|}{~} & \mc{O}_{32}=\langle \nabla^{\mu}\nabla^{\nu}u^{\rho}u^{\sigma}u_{\mu}u_{\nu}\rangle \langle u_{\rho}u_{\sigma}\rangle  & \mc{B}_{32} = \mc{Y} \circ \tr{1234|56} s_{13}s_{14}s_{15}s_{26}\\
		 & & & \multicolumn{2}{c|}{~} & \mc{O}_{33}=\langle \nabla^{\mu}u^{\nu}\nabla^{\rho}u^{\sigma}u_{\mu}u_{\rho}\rangle \langle u_{\nu}u_{\sigma}\rangle  & \mc{B}_{33} = \mc{Y} \circ \tr{1234|56} s_{13}s_{15}s_{24}s_{26}\\
		 & & & \multicolumn{2}{c|}{~} & \mc{O}_{34}=\langle \nabla^{\mu}u^{\nu}\nabla^{\rho}u^{\sigma}u_{\mu}u_{\nu}\rangle \langle u_{\rho}u_{\sigma}\rangle  & \mc{B}_{34} = \mc{Y} \circ \tr{1234|56} s_{13}s_{14}s_{25}s_{26}\\
		 & & & \multicolumn{2}{c|}{~} & \mc{O}_{35}=\langle \nabla^{\mu}u^{\nu}\nabla_{\mu}u^{\rho}u^{\sigma}u_{\sigma}\rangle \langle u_{\nu}u_{\rho}\rangle  & \mc{B}_{35} = \mc{Y} \circ \tr{1234|56} s_{12}s_{15}s_{26}s_{34}\\
		 & & & \multicolumn{2}{c|}{~} & \mc{O}_{36}=\langle \nabla^{\mu}u^{\nu}\nabla_{\mu}u^{\rho}u^{\sigma}u_{\rho}\rangle \langle u_{\nu}u_{\sigma}\rangle  & \mc{B}_{36} = \mc{Y} \circ \tr{1234|56} s_{12}s_{15}s_{24}s_{36}\\
		 & & & \multicolumn{2}{c|}{~} & \mc{O}_{37}=\langle \nabla^{\mu}u^{\nu}u^{\rho}\nabla_{\mu}u^{\sigma}u_{\rho}\rangle \langle u_{\nu}u_{\sigma}\rangle  & \mc{B}_{37} = \mc{Y} \circ \tr{1234|56} s_{13}s_{15}s_{24}s_{36}\\
		 & & & \multicolumn{2}{c|}{~} & \mc{O}_{38}=\langle \nabla^{\mu}u^{\nu}\nabla_{\mu}u^{\rho}u^{\sigma}u_{\nu}\rangle \langle u_{\rho}u_{\sigma}\rangle  & \mc{B}_{38} = \mc{Y} \circ \tr{1234|56} s_{12}s_{14}s_{25}s_{36}\\
		 & & & \multicolumn{2}{c|}{~} & \mc{O}_{39}=\langle \nabla^{\mu}u^{\nu}u^{\rho}\nabla_{\mu}u^{\sigma}u_{\nu}\rangle \langle u_{\rho}u_{\sigma}\rangle  & \mc{B}_{39} = \mc{Y} \circ \tr{1234|56} s_{13}s_{14}s_{25}s_{36}\\
		 & & & \multicolumn{2}{c|}{~} & \mc{O}_{40}=\langle \nabla^{\mu}\nabla^{\nu}u^{\rho}u_{\mu}u^{\nu}u^{\rho}\rangle \langle u^{\sigma}u_{\sigma}\rangle  & \mc{B}_{40} = \mc{Y} \circ \tr{1234|56} s_{12}s_{13}s_{14}s_{56}\\
        \hline
    \end{array}
\end{align*}
\end{table}

\begin{table}[htbp]
	\centering 
\begin{align*} \footnotesize
    \begin{array}{|c|c|c|c|c|l|l|}
        \hline
		\multicolumn{5}{|c|}{SU(N_f)} & \text{Operator Basis}  & \text{Amplitude Basis} \\
		\hline\hline
            & & \multicolumn{3}{c|}{~} & \mc{O}_{41}=\langle \nabla^{\mu}\nabla^{\nu}u^{\rho}u^{\sigma}u_{\sigma}u_{\mu}\rangle \langle u_{\nu}u_{\rho}\rangle   & \mc{B}_{41} = \mc{Y} \circ \tr{1234|56}s_{14}s_{15}s_{16}s_{23}   \\
		    & & \multicolumn{3}{c|}{~} & \mc{O}_{42}=\langle \nabla^{\mu}u^{\nu}\nabla^{\rho}u^{\sigma}u_{\rho}u_{\mu}\rangle \langle u_{\nu}u_{\sigma}\rangle  & \mc{B}_{42} = \mc{Y} \circ \tr{1234|56} s_{14}s_{15}s_{23}s_{26}    \\
		    & & \multicolumn{3}{c|}{~} & \mc{O}_{43}=\langle \nabla^{\mu}\nabla^{\nu}u^{\rho}u_{\mu}u^{\sigma}u_{\sigma}\rangle \langle u_{\nu}u_{\rho}\rangle  & \mc{B}_{43} = \mc{Y} \circ \tr{1234|56} s_{12}s_{15}s_{16}s_{34}    \\
		    & & \multicolumn{3}{c|}{\multirow{21}*{$SU(4)$}} & \mc{O}_{44}=\langle \nabla^{\mu}u^{\nu}u^{\rho}\nabla_{\mu}u^{\sigma}u_{\sigma}\rangle \langle u_{\nu}u_{\rho}\rangle  & \mc{B}_{44} = \mc{Y} \circ \tr{1234|56} s_{13}s_{15}s_{26}s_{34}   \\
		    & & \multicolumn{3}{c|}{~} & \mc{O}_{45}=\langle \nabla^{\mu}u^{\nu}u^{\rho}u^{\sigma}u_{\sigma}\rangle \langle \nabla_{\mu}u_{\nu}u_{\rho}\rangle   & \mc{B}_{45} = \mc{Y} \circ \tr{1234|56} s_{15}^2s_{26}s_{34}   \\
		    & & \multicolumn{3}{c|}{~} & \mc{O}_{46}=\langle \nabla^{\mu}\nabla^{\nu}u^{\rho}u_{\mu}u^{\sigma}u_{\nu}\rangle \langle u_{\rho}u_{\sigma}\rangle   & \mc{B}_{46} = \mc{Y} \circ \tr{1234|56} s_{12}s_{14}s_{15}s_{36}   \\
		    & & \multicolumn{3}{c|}{~} & \mc{O}_{47}=\langle \nabla^{\mu}u^{\nu}u^{\rho}\nabla_{\rho}u^{\sigma}u_{\mu}\rangle \langle u_{\nu}u_{\sigma}\rangle   & \mc{B}_{47} = \mc{Y} \circ \tr{1234|56} s_{14}s_{15}s_{23}s_{36}   \\
			& & \multicolumn{3}{c|}{~} & \mc{O}_{48}=\langle \nabla^{\mu}u^{\nu}u^{\rho}u^{\sigma}\nabla_{\rho}u_{\mu}\rangle \langle u_{\nu}u_{\sigma}\rangle   & \mc{B}_{48} = \mc{Y} \circ \tr{1234|56} s_{14}s_{15}s_{24}s_{36}   \\
			& & \multicolumn{3}{c|}{~} & \mc{O}_{49}=\langle \nabla^{\mu}u^{\nu}u^{\rho}u^{\sigma}\nabla_{\mu}u_{\nu}\rangle \langle u_{\rho}u_{\sigma}\rangle   & \mc{B}_{49} = \mc{Y} \circ \tr{1234|56} s_{14}^2s_{25}s_{36}  \\
			& & \multicolumn{3}{c|}{~} & \mc{O}_{50}=\langle \nabla^{\mu}\nabla^{\nu}u^{\rho}u_{\mu}u_{\nu}u^{\sigma}\rangle \langle u_{\rho}u_{\sigma}\rangle   & \mc{B}_{50} = \mc{Y} \circ \tr{1234|56} s_{12}s_{13}s_{15}s_{46}   \\
			& & \multicolumn{3}{c|}{~} & \mc{O}_{51}=\langle \nabla^{\mu}u^{\nu}\nabla_{\mu}u^{\rho}u_{\rho}u^{\sigma}\rangle \langle u^{\nu}u_{\sigma}\rangle   & \mc{B}_{51} = \mc{Y} \circ \tr{1234|56} s_{12}s_{15}s_{23}s_{46}   \\
			& & \multicolumn{3}{c|}{~} & \mc{O}_{52}=\langle \nabla^{\mu}u^{\nu}u^{\rho}\nabla_{\mu}u_{\nu}u^{\sigma}\rangle \langle u_{\rho}u_{\sigma}\rangle   & \mc{B}_{52} = \mc{Y} \circ \tr{1234|56} s_{13}^2s_{25}s_{46}   \\
			& & \multicolumn{3}{c|}{~} & \mc{O}_{53}=\langle u^{\mu}u^{\nu}u^{\rho}u^{\sigma}\rangle \langle \nabla_{\mu}\nabla_{\nu}u_{\rho}u_{\sigma}\rangle   & \mc{B}_{53} = \mc{Y} \circ \tr{1234|56} s_{15}s_{25}s_{35}s_{46}   \\
			& & \multicolumn{3}{c|}{~} & \mc{O}_{54}=\langle \nabla^{\mu}u^{\nu}\nabla_{\mu}u^{\rho}u_{\nu}u_{\rho}\rangle \langle u^{\sigma}u_{\sigma}\rangle   & \mc{B}_{54} = \mc{Y} \circ \tr{1234|56} s_{12}s_{13}s_{24}s_{56}   \\
			 & & \multicolumn{3}{c|}{~} & \mc{O}_{55}=\langle \nabla^{\mu}\nabla^{\nu}u^{\rho}u^{\sigma}u_{\sigma}\rangle \langle u_{\mu}u_{\nu}u_{\rho}\rangle   & \mc{B}_{55} = \mc{Y} \circ \tr{123|456} s_{14}s_{15}s_{16}s_{23}   \\
			 & & \multicolumn{3}{c|}{~} & \mc{O}_{56}=\langle \nabla^{\mu}\nabla^{\nu}u^{\rho}u^{\sigma}u_{\mu}\rangle \langle u_{\nu}u_{\rho}u_{\sigma}\rangle   & \mc{B}_{56} = \mc{Y} \circ \tr{123|456} s_{13}s_{14}s_{15}s_{26}    \\
			 & & \multicolumn{3}{c|}{~} & \mc{O}_{57}=\langle \nabla^{\mu}u^{\nu}\nabla^{\rho}u^{\sigma}u_{\rho}\rangle \langle u_{\mu}u_{\nu}u_{\sigma}\rangle   & \mc{B}_{57} = \mc{Y} \circ \tr{123|456} s_{14}s_{15}s_{23}s_{26}    \\
			 & & \multicolumn{3}{c|}{~} & \mc{O}_{58}=\langle \nabla^{\mu}u^{\nu}\nabla^{\rho}u^{\sigma}u_{\mu}\rangle \langle u_{\nu}u_{\rho}u_{\sigma}\rangle   & \mc{B}_{58} = \mc{Y} \circ \tr{123|456} s_{13}s_{14}s_{25}s_{26}    \\
			 & & \multicolumn{3}{c|}{~} & \mc{O}_{59}=\langle \nabla^{\mu}u^{\nu}\nabla_{\mu}u^{\rho}u^{\sigma}\rangle \langle u_{\nu}u_{\sigma}u_{\rho}\rangle   & \mc{B}_{59} = \mc{Y} \circ \tr{123|456} s_{12}s_{14}s_{26}s_{35}    \\
			 & & \multicolumn{3}{c|}{~} & \mc{O}_{60}=\langle \nabla^{\mu}u^{\nu}u^{\rho}u^{\sigma}\rangle \langle \nabla_{\mu}u_{\nu}u_{\sigma}u_{\rho}\rangle   & \mc{B}_{60} = \mc{Y} \circ \tr{123|456} s_{14}^2s_{26}s_{35}    \\
			 & & \multicolumn{3}{c|}{~} & \mc{O}_{61}=\langle \nabla^{\mu}u^{\nu}\nabla_{\mu}u^{\rho}u^{\sigma}\rangle \langle u_{\nu}u_{\rho}u_{\sigma}\rangle   & \mc{B}_{61} = \mc{Y} \circ \tr{123|456} s_{12}s_{14}s_{25}s_{36}    \\
			 & & \multicolumn{3}{c|}{~} & \mc{O}_{62}=\langle \nabla^{\mu}u^{\nu}u^{\rho}u^{\sigma}\rangle \langle \nabla_{\mu}u_{\nu}u_{\rho}u_{\sigma}\rangle   & \mc{B}_{62} = \mc{Y} \circ \tr{123|456} s_{14}^2s_{25}s_{36}    \\
			 & & \multicolumn{3}{c|}{~} & \mc{O}_{63}=\langle \nabla^{\mu}\nabla^{\nu}u^{\rho}u_{\mu}u_{\nu}\rangle \langle u^{\sigma}u_{\rho}u_{\sigma}\rangle   & \mc{B}_{63} = \mc{Y} \circ \tr{123|456} s_{12}s_{13}s_{15}s_{46}    \\
			 & & \multicolumn{3}{c|}{~} & \mc{O}_{64}=\langle \nabla^{\mu}u^{\nu}\nabla_{\mu}u^{\rho}u_{\rho}\rangle \langle u^{\sigma}u_{\nu}u_{\sigma}\rangle   & \mc{B}_{64} = \mc{Y} \circ \tr{123|456} s_{12}s_{15}s_{23}s_{46}    \\
			 & & \multicolumn{3}{c|}{~} & \mc{O}_{65}=\langle \nabla^{\mu}u^{\nu}\nabla_{\mu}u^{\rho}u_{\nu}\rangle \langle u^{\sigma}u_{\rho}u_{\sigma}\rangle   & \mc{B}_{65} = \mc{Y} \circ \tr{123|456} s_{12}s_{13}s_{25}s_{46}    \\
			 & & \multicolumn{3}{c|}{~} & \mc{O}_{66}=\langle \nabla^{\mu}u^{\nu}u^{\rho}u_{\mu}\rangle \langle \nabla_{\nu}u^{\sigma}u_{\rho}u_{\sigma}\rangle   & \mc{B}_{66} = \mc{Y} \circ \tr{123|456} s_{13}s_{14}s_{25}s_{46}    \\
			 & & \multicolumn{3}{c|}{~} & \mc{O}_{67}=\langle \nabla^{\mu}u^{\nu}\nabla^{\rho}u^{\sigma}\rangle \langle u_{\mu}u_{\nu}\rangle \langle u_{\rho}u_{\sigma}\rangle   & \mc{B}_{67} = \mc{Y} \circ \tr{12|34|56} s_{13}s_{14}s_{25}s_{26}   \\
			 & & \multicolumn{3}{c|}{~} & \mc{O}_{68}=\langle \nabla^{\mu}u^{\nu}\nabla_{\mu}u^{\rho}\rangle \langle u^{\sigma}u_{\nu}\rangle \langle u_{\rho}u_{\sigma}\rangle   & \mc{B}_{68} = \mc{Y} \circ \tr{12|34|56} s_{12}s_{14}s_{25}s_{36}   \\
			 \cline{3-7}  
			 & \multicolumn{4}{c|}{\multirow{6}*{$SU(5)$}} & \mc{O}_{69}=\langle \nabla^{\mu}\nabla^{\nu}u^{\rho}u_{\mu}u^{\sigma}\rangle \langle u_{\nu}u_{\rho}u_{\sigma}\rangle   & \mc{B}_{69} = \mc{Y} \circ \tr{123|456} s_{12}s_{14}s_{15}s_{36}   \\
			 & \multicolumn{4}{c|}{~} & \mc{O}_{70}=\langle \nabla^{\mu}\nabla^{\nu}u^{\rho}u^{\sigma}\rangle \langle u_{\mu}u_{\nu}\rangle \langle u_{\rho}u_{\sigma}\rangle   & \mc{B}_{70} = \mc{Y} \circ \tr{12|34|56} s_{13}s_{14}s_{15}s_{26}   \\
			 & \multicolumn{4}{c|}{~} & \mc{O}_{71}=\langle \nabla^{\mu}u^{\nu}\nabla^{\rho}u^{\sigma}\rangle \langle u_{\mu}u_{\rho}\rangle \langle u_{\nu}u_{\sigma}\rangle   & \mc{B}_{71} = \mc{Y} \circ \tr{12|34|56} s_{13}s_{15}s_{24}s_{26}   \\
			 & \multicolumn{4}{c|}{~} & \mc{O}_{72}=\langle \nabla^{\mu}u^{\nu}\nabla_{\mu}u^{\rho}\rangle \langle u^{\sigma}u_{\sigma}\rangle \langle u_{\nu}u_{\rho}\rangle   & \mc{B}_{72} = \mc{Y} \circ \tr{12|34|56} s_{12}s_{15}s_{26}s_{34}   \\
			 & \multicolumn{4}{c|}{~} & \mc{O}_{73}=\langle \nabla^{\mu}\nabla^{\nu}u^{\rho}u_{\mu}\rangle \langle u^{\sigma}u_{\nu}\rangle \langle u_{\rho}u_{\sigma}\rangle   & \mc{B}_{73} = \mc{Y} \circ \tr{12|34|56} s_{12}s_{14}s_{15}s_{36}   \\
			 & \multicolumn{4}{c|}{~} & \mc{O}_{74}=\langle \nabla^{\mu}u^{\nu}u^{\rho}\rangle \langle \nabla_{\mu}u^{\sigma}u_{\nu}\rangle \langle u_{\rho}u_{\sigma}\rangle   & \mc{B}_{74} = \mc{Y} \circ \tr{12|34|56} s_{13}s_{14}s_{25}s_{36}   \\
			 \cline{2-7}  
			 \multicolumn{5}{|c|}{\multirow{2}*{$SU(N_f \ge 6)$}} & \mc{O}_{75}=\langle \nabla^{\mu}\nabla^{\nu}u^{\rho}u_{\mu}\rangle \langle u^{\sigma}u_{\sigma}\rangle \langle u_{\nu}u_{\rho}\rangle    & \mc{B}_{75} = \mc{Y} \circ \tr{12|34|56} s_{12}s_{15}s_{16}s_{34}  \\
			 \multicolumn{5}{|c|}{~} & \mc{O}_{76}=\langle \nabla^{\mu}u^{\nu}u^{\rho}\rangle \langle \nabla_{\mu}u^{\sigma}u_{\rho}\rangle \langle u_{\nu}u_{\sigma}\rangle    & \mc{B}_{76} = \mc{Y} \circ \tr{12|34|56} s_{13}s_{15}s_{24}s_{36}  \\   
		\hline
	\end{array}
\end{align*}
\caption{A complete set of 6-Goldstone parity-even operators in $p^8$ order. The number of independent monomials for $SU(2),SU(3),SU(4),SU(5)$ and $SU(N_f\ge 6)$ flavour groups are $8,40,68,74$ and $76$ respectively, agreeing with \cite{Graf:2020yxt}.} \label{tab:even682}
\end{table}

\begin{table}[htbp]
	\centering
\begin{align*}
    \begin{array}{|c|c|c|c|l|l|}
    \hline
		\multicolumn{4}{|c|}{SU(N_f)} & \text{Operator Basis}  & \text{Amplitude Basis} \\
    \hline\hline
         & & & \multirow{2}*{$SU(2)$} & \mc{O}_1=\langle \nabla^{\mu}u^{\nu}\nabla_{\mu}u^{\rho}u_{\nu}u^{\sigma}u^{\eta}u^{\lambda}\rangle\epsilon_{\rho\sigma\eta\lambda} & \mc{B}_1 = \mc{Y} \circ \tr{123456} s_{12}s_{13}\epsilon(2,4,5,6)\\
         & & & & \mc{O}_2=\langle \nabla^{\mu}u^{\rho}\nabla^{\sigma}u^{\nu}u_{\nu}u_{\mu}u^{\eta}u^{\lambda}\rangle\epsilon_{\rho\sigma\eta\lambda} & \mc{B}_2 = \mc{Y} \circ \tr{123456} s_{14}s_{23}\epsilon(1,2,5,6)\\
         \cline{4-6}
         & & \multicolumn{2}{c|}{\multirow{18}*{$SU(3)$}} & \mc{O}_3=\langle \nabla^{\mu}\nabla^{\nu}u^{\rho}u^{\sigma}u_{\mu}u^{\eta}u_{\nu}u^{\lambda} \rangle\epsilon_{\rho\sigma\eta\lambda} & \mc{B}_3 = \mc{Y} \circ \tr{123456} s_{13}s_{15}\epsilon(1,2,4,6) \\
         & & \multicolumn{2}{c|}{~} & \mc{O}_4=\langle \nabla^{\mu}\nabla^{\nu}u^{\rho}u^{\sigma}u_{\mu}u_{\nu}u^{\eta}u^{\lambda} \rangle\epsilon_{\rho\sigma\eta\lambda}  & \mc{B}_4 = \mc{Y} \circ \tr{123456} s_{13}s_{14}\epsilon(1,2,5,6)\\
         & & \multicolumn{2}{c|}{~} & \mc{O}_5=\langle \nabla^{\mu}\nabla^{\nu}u^{\rho}u_{\mu}u^{\sigma}u^{\eta}u_{\nu}u^{\lambda} \rangle\epsilon_{\rho\sigma\eta\lambda}  & \mc{B}_5 = \mc{Y} \circ \tr{123456} s_{12}s_{15}\epsilon(1,3,4,6)\\
         & & \multicolumn{2}{c|}{~} & \mc{O}_6=\langle \nabla^{\mu}\nabla^{\nu}u^{\rho}u_{\mu}u^{\sigma}u_{\nu}u^{\eta}u^{\lambda} \rangle\epsilon_{\rho\sigma\eta\lambda}  & \mc{B}_6 = \mc{Y} \circ \tr{123456} s_{12}s_{14}\epsilon(1,3,5,6)\\
         & & \multicolumn{2}{c|}{~} & \mc{O}_7=\langle \nabla^{\mu}u^{\rho}u^{\nu}\nabla_{\mu}u^{\sigma}u_{\nu}u^{\eta}u^{\lambda} \rangle\epsilon_{\rho\sigma\eta\lambda}  & \mc{B}_7 = \mc{Y} \circ \tr{123456} s_{13}s_{24}\epsilon(1,3,5,6)\\
		 & & \multicolumn{2}{c|}{~} & \mc{O}_8=\langle \nabla^{\mu}\nabla^{\nu}u^{\rho}u_{\mu}u_{\nu}u^{\sigma}u^{\eta}u^{\lambda} \rangle\epsilon_{\rho\sigma\eta\lambda}  & \mc{B}_8 = \mc{Y} \circ \tr{123456} s_{12}s_{13}\epsilon(1,4,5,6)\\
		 & & \multicolumn{2}{c|}{~} & \mc{O}_9=\langle \nabla^{\mu}u^{\nu}\nabla_{\mu}u^{\rho}u^{\sigma}u^{\eta}u_{\nu}u^{\lambda} \rangle\epsilon_{\rho\sigma\eta\lambda}  & \mc{B}_9 = \mc{Y} \circ \tr{123456} s_{12}s_{15}\epsilon(2,3,4,6)\\
		 & & \multicolumn{2}{c|}{~} & \mc{O}_{10}=\langle \nabla^{\mu}u^{\nu}u^{\rho}\nabla_{\mu}u^{\sigma}u^{\eta}u_{\nu}u^{\lambda} \rangle\epsilon_{\rho\sigma\eta\lambda}  & \mc{B}_{10} = \mc{Y} \circ \tr{123456} s_{13}s_{15}\epsilon(2,3,4,6)\\
		 & & \multicolumn{2}{c|}{~} & \mc{O}_{11}=\langle \nabla^{\mu}u^{\nu}\nabla_{\mu}u^{\rho}u^{\sigma}u_{\nu}u^{\eta}u^{\lambda} \rangle\epsilon_{\rho\sigma\eta\lambda}  & \mc{B}_{11} = \mc{Y} \circ \tr{123456} s_{12}s_{14}\epsilon(2,3,5,6)\\
		 & & \multicolumn{2}{c|}{~} & \mc{O}_{12}=\langle \nabla^{\mu}u^{\nu}u^{\rho}\nabla_{\mu}u^{\sigma}u_{\nu}u^{\eta}u^{\lambda} \rangle\epsilon_{\rho\sigma\eta\lambda}  & \mc{B}_{12} = \mc{Y} \circ \tr{123456} s_{13}s_{14}\epsilon(2,3,5,6)\\
		 & & \multicolumn{2}{c|}{~} & \mc{O}_{13}=\langle \nabla^{\mu}u^{\nu}u^{\rho}u^{\sigma}\nabla_{\mu}u_{\nu}u^{\eta}u^{\lambda} \rangle\epsilon_{\rho\sigma\eta\lambda}  & \mc{B}_{13} = \mc{Y} \circ \tr{123456} s_{14}^2\epsilon(2,3,5,6)\\
		 & & \multicolumn{2}{c|}{~} & \mc{O}_{14}=\langle \nabla^{\mu}u^{\nu}u^{\rho}\nabla_{\mu}u_{\nu}u^{\sigma}u^{\eta}u^{\lambda} \rangle\epsilon_{\rho\sigma\eta\lambda}  & \mc{B}_{14} = \mc{Y} \circ \tr{123456} s_{13}^2\epsilon(2,4,5,6)\\
		 & & \multicolumn{2}{c|}{~} & \mc{O}_{15}=\langle \nabla^{\mu}u^{\nu}\nabla_{\mu}u_{\nu}u^{\rho}u^{\sigma}u^{\eta}u^{\lambda} \rangle\epsilon_{\rho\sigma\eta\lambda}  & \mc{B}_{15} = \mc{Y} \circ \tr{123456} s_{12}^2\epsilon(3,4,5,6)\\
		 & & \multicolumn{2}{c|}{~} & \mc{O}_{16}=\langle \nabla^{\mu}\nabla^{\nu}u^{\rho}u^{\sigma}u^{\eta}u_{\mu}u_{\nu}u^{\lambda} \rangle\epsilon_{\rho\sigma\eta\lambda}  & \mc{B}_{16} = \mc{Y} \circ \tr{123456} s_{14}s_{15}\epsilon(1,2,3,6)\\
		 & & \multicolumn{2}{c|}{~} & \mc{O}_{17}=\langle \nabla^{\mu}u^{\rho}\nabla^{\nu}u^{\sigma}u_{\mu}u^{\eta} \rangle\langle u_{\nu}u^{\lambda} \rangle\epsilon_{\rho\sigma\eta\lambda}  & \mc{B}_{17} = \mc{Y} \circ \tr{1234|56} s_{13}s_{25}\epsilon(1,2,4,6)\\
		 & & \multicolumn{2}{c|}{~} & \mc{O}_{18}=\langle \nabla^{\mu}\nabla^{\nu}u^{\rho}u_{\mu}u^{\sigma}u^{\eta} \rangle\langle u_{\nu}u^{\lambda} \rangle\epsilon_{\rho\sigma\eta\lambda}  & \mc{B}_{18} = \mc{Y} \circ \tr{1234|56} s_{12}s_{15}\epsilon(1,3,4,6)\\
		 & & \multicolumn{2}{c|}{~} & \mc{O}_{19}=\langle \nabla^{\mu}u^{\nu}\nabla_{\mu}u^{\rho}u^{\sigma}u^{\eta} \rangle\langle u_{\nu}u^{\lambda} \rangle\epsilon_{\rho\sigma\eta\lambda}  & \mc{B}_{19} = \mc{Y} \circ \tr{1234|56} s_{12}s_{15}\epsilon(2,3,4,6)\\
		 & & \multicolumn{2}{c|}{~} & \mc{O}_{20}=\langle \nabla^{\mu}u^{\nu}u^{\rho}\nabla_{\mu}u^{\sigma}u^{\eta} \rangle\langle u_{\nu}u^{\lambda} \rangle\epsilon_{\rho\sigma\eta\lambda}  & \mc{B}_{20} = \mc{Y} \circ \tr{1234|56} s_{13}s_{15}\epsilon(2,3,4,6)\\
         \cline{3-6}
         & \multicolumn{3}{c|}{\multirow{13}*{$SU(4)$}} & \mc{O}_{21}=\langle \nabla^{\mu}\nabla^{\nu}u^{\rho}u^{\sigma}u^{\eta}u_{\mu}\rangle \langle u_{\nu}u^{\lambda}\rangle\epsilon_{\rho\sigma\eta\lambda}   & \mc{B}_{21} = \mc{Y} \circ \tr{1234|56}s_{14}s_{15}\epsilon(1,2,3,6)   \\
		 & \multicolumn{3}{c|}{~} & \mc{O}_{22}=\langle \nabla^{\mu}u^{\rho}\nabla^{\nu}u^{\sigma}u^{\eta}u_{\mu}\rangle \langle u_{\nu}u^{\lambda}\rangle\epsilon_{\rho\sigma\eta\lambda}  & \mc{B}_{22} = \mc{Y} \circ \tr{1234|56} s_{14}s_{25}\epsilon(1,2,3,6)    \\
		 & \multicolumn{3}{c|}{~} & \mc{O}_{23}=\langle \nabla^{\mu}\nabla^{\nu}u^{\rho}u^{\sigma}u_{\mu}u^{\eta}\rangle \langle u_{\nu}u^{\lambda}\rangle\epsilon_{\rho\sigma\eta\lambda}  & \mc{B}_{23} = \mc{Y} \circ \tr{1234|56} s_{13}s_{15}\epsilon(1,2,4,6)    \\
		 & \multicolumn{3}{c|}{~} & \mc{O}_{24}=\langle \nabla^{\mu}u^{\rho}u^{\nu}\nabla_{\mu}u^{\sigma}u^{\eta}\rangle \langle u_{\nu}u^{\lambda}\rangle\epsilon_{\rho\sigma\eta\lambda}  & \mc{B}_{24} = \mc{Y} \circ \tr{1234|56} s_{13}s_{25}\epsilon(1,3,4,6)   \\
		 & \multicolumn{3}{c|}{~} & \mc{O}_{25}=\langle \nabla^{\mu}u^{\rho}u^{\nu}u^{\sigma}\nabla_{\mu}u^{\eta}\rangle \langle u_{\nu}u^{\lambda}\rangle\epsilon_{\rho\sigma\eta\lambda}   & \mc{B}_{25} = \mc{Y} \circ \tr{1234|56} s_{14}s_{25}\epsilon(1,3,4,6)   \\
		 & \multicolumn{3}{c|}{~} & \mc{O}_{26}=\langle \nabla^{\mu}u^{\nu}u^{\rho}u^{\sigma}\nabla_{\mu}u^{\eta}\rangle \langle u_{\nu}u^{\lambda}\rangle\epsilon_{\rho\sigma\eta\lambda}   & \mc{B}_{26} = \mc{Y} \circ \tr{1234|56} s_{14}s_{15}\epsilon(2,3,4,6)   \\
		 & \multicolumn{3}{c|}{~} & \mc{O}_{27}=\langle u^{\mu}\nabla^{\nu}u^{\rho}u^{\sigma}\nabla_{\mu}u^{\eta}\rangle \langle u_{\nu}u^{\lambda}\rangle\epsilon_{\rho\sigma\eta\lambda}   & \mc{B}_{27} = \mc{Y} \circ \tr{1234|56} s_{14}s_{25}\epsilon(2,3,4,6)   \\
		 & \multicolumn{3}{c|}{~} & \mc{O}_{28}=\langle \nabla^{\mu}u^{\rho}u^{\nu}u^{\sigma}\nabla_{\mu} \rangle\langle u^{\eta}u_{\nu}u^{\lambda}\rangle\epsilon_{\rho\sigma\eta\lambda}   & \mc{B}_{28} = \mc{Y} \circ \tr{123|456} s_{14}s_{25}\epsilon(1,3,4,6)   \\
		 & \multicolumn{3}{c|}{~} & \mc{O}_{29}=\langle \nabla^{\mu}\nabla^{\nu}u^{\rho}u_{\mu}u_{\nu} \rangle\langle u^{\sigma}u^{\eta}u^{\lambda}\rangle\epsilon_{\rho\sigma\eta\lambda}   & \mc{B}_{29} = \mc{Y} \circ \tr{123|456} s_{12}s_{13}\epsilon(1,4,5,6)  \\
		 & \multicolumn{3}{c|}{~} & \mc{O}_{30}=\langle \nabla^{\mu}u^{\nu}u^{\rho}u^{\sigma}\nabla_{\mu} \rangle\langle u_{\nu}u^{\eta}u^{\lambda}\rangle\epsilon_{\rho\sigma\eta\lambda}   & \mc{B}_{30} = \mc{Y} \circ \tr{123|456} s_{14}^2\epsilon(2,3,5,6)   \\
		 & \multicolumn{3}{c|}{~} & \mc{O}_{31}=\langle \nabla^{\mu}u^{\nu}\nabla_{\mu}u^{\rho}u_{\nu} \rangle\langle u^{\sigma}u^{\eta}u^{\lambda}\rangle\epsilon_{\rho\sigma\eta\lambda}   & \mc{B}_{31} = \mc{Y} \circ \tr{123|456} s_{12}s_{13}\epsilon(2,4,5,6)   \\
		 & \multicolumn{3}{c|}{~} & \mc{O}_{32}=\langle \nabla^{\mu}u^{\nu}\nabla_{\mu}u_{\nu}u^{\rho} \rangle\langle u^{\sigma}u^{\eta}u^{\lambda}\rangle\epsilon_{\rho\sigma\eta\lambda}   & \mc{B}_{32} = \mc{Y} \circ \tr{123|456} s_{12}^2\epsilon(3,4,5,6)   \\
		 & \multicolumn{3}{c|}{~} & \mc{O}_{33}=\langle \nabla^{\mu}u^{\nu}u^{\rho}\nabla_{\mu} \rangle\langle u^{\sigma}u^{\eta} \rangle\langle u_{\nu}u^{\lambda} \rangle\epsilon_{\rho\sigma\eta\lambda}   & \mc{B}_{33} = \mc{Y} \circ \tr{12|34|56} s_{13}s_{15}\epsilon(2,3,4,6)   \\
		 \cline{2-6}  
		 \multicolumn{4}{|c|}{\multirow{2}*{$SU(N_f \ge 5)$}} & \mc{O}_{34}=\langle \nabla^{\mu}\nabla^{\nu}u^{\rho}u^{\sigma}u^{\eta} \rangle\langle u_{\mu}u_{\nu}u^{\lambda}\rangle\epsilon_{\rho\sigma\eta\lambda}    & \mc{B}_{34} = \mc{Y} \circ \tr{123|456} s_{14}s_{15}\epsilon(1,2,3,6)  \\
		 \multicolumn{4}{|c|}{~} & \mc{O}_{35}=\langle \nabla^{\mu}u^{\rho}\nabla^{\nu}u^{\sigma}\rangle \langle u_{\mu}u^{\eta}\rangle \langle u_{\nu}u^{\lambda}\rangle\epsilon_{\rho\sigma\eta\lambda}    & \mc{B}_{35} = \mc{Y} \circ \tr{12|34|56} s_{13}s_{25}\epsilon(1,2,4,6)  \\
    \hline
    \end{array}
\end{align*}
\caption{A complete set of 6-Goldstone parity-odd operators in $p^8$ order. The number of independent monomials for $SU(2),SU(3),SU(4)$ and $SU(N_f\ge 5)$ flavour groups are $2,20,33$ and $35$ respectively, agreeing with \cite{Graf:2020yxt}.}
\label{tab:odd68}
\end{table}

\newpage


\bibliography{nlsm}

\providecommand{\href}[2]{#2}\begingroup\raggedright\begin{thebibliography}{10}

\bibitem{Weinberg:1968de}
S.~Weinberg, ``{Nonlinear realizations of chiral symmetry},''
  \href{http://dx.doi.org/10.1103/PhysRev.166.1568}{{\em Phys. Rev.} {\bfseries
  166} (1968) 1568--1577}.

\bibitem{Weinberg:1978kz}
S.~Weinberg, ``{Phenomenological Lagrangians},''
  \href{http://dx.doi.org/10.1016/0378-4371(79)90223-1}{{\em Physica A}
  {\bfseries 96} no.~1-2, (1979) 327--340}.

\bibitem{Gasser:1983yg}
J.~Gasser and H.~Leutwyler, ``{Chiral Perturbation Theory to One Loop},''
  \href{http://dx.doi.org/10.1016/0003-4916(84)90242-2}{{\em Annals Phys.}
  {\bfseries 158} (1984) 142}.

\bibitem{Fearing:1994ga}
H.~Fearing and S.~Scherer, ``{Extension of the chiral perturbation theory meson
  Lagrangian to order p(6)},''
  \href{http://dx.doi.org/10.1103/PhysRevD.53.315}{{\em Phys. Rev. D}
  {\bfseries 53} (1996) 315--348},
  \href{http://arxiv.org/abs/hep-ph/9408346}{{\ttfamily arXiv:hep-ph/9408346}}.

\bibitem{Bijnens:1999sh}
J.~Bijnens, G.~Colangelo, and G.~Ecker, ``{The Mesonic chiral Lagrangian of
  order p**6},'' \href{http://dx.doi.org/10.1088/1126-6708/1999/02/020}{{\em
  JHEP} {\bfseries 02} (1999) 020},
  \href{http://arxiv.org/abs/hep-ph/9902437}{{\ttfamily arXiv:hep-ph/9902437}}.

\bibitem{Bijnens:2018lez}
J.~Bijnens, N.~Hermansson-Truedsson, and S.~Wang, ``{The order p$^{8}$ mesonic
  chiral Lagrangian},'' \href{http://dx.doi.org/10.1007/JHEP01(2019)102}{{\em
  JHEP} {\bfseries 01} (2019) 102},
  \href{http://arxiv.org/abs/1810.06834}{{\ttfamily arXiv:1810.06834
  [hep-ph]}}.

\bibitem{Weinberg:1979sa}
S.~Weinberg, ``{Baryon and Lepton Nonconserving Processes},''
  \href{http://dx.doi.org/10.1103/PhysRevLett.43.1566}{{\em Phys. Rev. Lett.}
  {\bfseries 43} (1979) 1566--1570}.

\bibitem{Buchmuller:1985jz}
W.~Buchmuller and D.~Wyler, ``{Effective Lagrangian Analysis of New
  Interactions and Flavor Conservation},''
  \href{http://dx.doi.org/10.1016/0550-3213(86)90262-2}{{\em Nucl. Phys. B}
  {\bfseries 268} (1986) 621--653}.

\bibitem{Grzadkowski:2010es}
B.~Grzadkowski, M.~Iskrzynski, M.~Misiak, and J.~Rosiek, ``{Dimension-Six Terms
  in the Standard Model Lagrangian},''
  \href{http://dx.doi.org/10.1007/JHEP10(2010)085}{{\em JHEP} {\bfseries 10}
  (2010) 085}, \href{http://arxiv.org/abs/1008.4884}{{\ttfamily arXiv:1008.4884
  [hep-ph]}}.

\bibitem{Adler:1964um}
S.~L. Adler, ``{Consistency conditions on the strong interactions implied by a
  partially conserved axial vector current},''
  \href{http://dx.doi.org/10.1103/PhysRev.137.B1022}{{\em Phys. Rev.}
  {\bfseries 137} (1965) B1022--B1033}.

\bibitem{Low:2014nga}
I.~Low, ``{Adler\textquoteright{}s zero and effective Lagrangians for
  nonlinearly realized symmetry},''
  \href{http://dx.doi.org/10.1103/PhysRevD.91.105017}{{\em Phys. Rev. D}
  {\bfseries 91} no.~10, (2015) 105017},
  \href{http://arxiv.org/abs/1412.2145}{{\ttfamily arXiv:1412.2145 [hep-th]}}.

\bibitem{Low:2014oga}
I.~Low, ``{Minimally symmetric Higgs boson},''
  \href{http://dx.doi.org/10.1103/PhysRevD.91.116005}{{\em Phys. Rev.}
  {\bfseries D91} no.~11, (2015) 116005},
\href{http://arxiv.org/abs/1412.2146}{{\ttfamily arXiv:1412.2146 [hep-ph]}}.

\bibitem{Cheung:2014dqa}
C.~Cheung, K.~Kampf, J.~Novotny, and J.~Trnka, ``{Effective Field Theories from
  Soft Limits of Scattering Amplitudes},''
  \href{http://dx.doi.org/10.1103/PhysRevLett.114.221602}{{\em Phys. Rev.
  Lett.} {\bfseries 114} no.~22, (2015) 221602},
  \href{http://arxiv.org/abs/1412.4095}{{\ttfamily arXiv:1412.4095 [hep-th]}}.

\bibitem{Cheung:2015ota}
C.~Cheung, K.~Kampf, J.~Novotny, C.-H. Shen, and J.~Trnka, ``{On-Shell
  Recursion Relations for Effective Field Theories},''
  \href{http://dx.doi.org/10.1103/PhysRevLett.116.041601}{{\em Phys. Rev.
  Lett.} {\bfseries 116} no.~4, (2016) 041601},
  \href{http://arxiv.org/abs/1509.03309}{{\ttfamily arXiv:1509.03309
  [hep-th]}}.

\bibitem{Low:2019ynd}
I.~Low and Z.~Yin, ``{Soft Bootstrap and Effective Field Theories},''
  \href{http://dx.doi.org/10.1007/JHEP11(2019)078}{{\em JHEP} {\bfseries 11}
  (2019) 078},
\href{http://arxiv.org/abs/1904.12859}{{\ttfamily arXiv:1904.12859 [hep-th]}}.

\bibitem{Dai:2020cpk}
L.~Dai, I.~Low, T.~Mehen, and A.~Mohapatra, ``{Operator Counting and Soft
  Blocks in Chiral Perturbation Theory},''
  \href{http://dx.doi.org/10.1103/PhysRevD.102.116011}{{\em Phys. Rev. D}
  {\bfseries 102} (2020) 116011},
  \href{http://arxiv.org/abs/2009.01819}{{\ttfamily arXiv:2009.01819
  [hep-ph]}}.

\bibitem{Liu:2018vel}
D.~Liu, I.~Low, and Z.~Yin, ``{Universal Imprints of a Pseudo-Nambu-Goldstone
  Higgs Boson},'' \href{http://dx.doi.org/10.1103/PhysRevLett.121.261802}{{\em
  Phys. Rev. Lett.} {\bfseries 121} no.~26, (2018) 261802},
  \href{http://arxiv.org/abs/1805.00489}{{\ttfamily arXiv:1805.00489
  [hep-ph]}}.

\bibitem{Liu:2018qtb}
D.~Liu, I.~Low, and Z.~Yin, ``{Universal Relations in Composite Higgs
  Models},'' \href{http://dx.doi.org/10.1007/JHEP05(2019)170}{{\em JHEP}
  {\bfseries 05} (2019) 170}, \href{http://arxiv.org/abs/1809.09126}{{\ttfamily
  arXiv:1809.09126 [hep-ph]}}.

\bibitem{Graf:2020yxt}
L.~Graf, B.~Henning, X.~Lu, T.~Melia, and H.~Murayama, ``{2, 12, 117, 1959,
  45171, 1170086, \textellipsis{}: a Hilbert series for the QCD chiral
  Lagrangian},'' \href{http://dx.doi.org/10.1007/JHEP01(2021)142}{{\em JHEP}
  {\bfseries 01} (2021) 142}, \href{http://arxiv.org/abs/2009.01239}{{\ttfamily
  arXiv:2009.01239 [hep-ph]}}.

\bibitem{Henning:2015daa}
B.~Henning, X.~Lu, T.~Melia, and H.~Murayama, ``{Hilbert series and operator
  bases with derivatives in effective field theories},''
  \href{http://dx.doi.org/10.1007/s00220-015-2518-2}{{\em Commun. Math. Phys.}
  {\bfseries 347} no.~2, (2016) 363--388},
  \href{http://arxiv.org/abs/1507.07240}{{\ttfamily arXiv:1507.07240
  [hep-th]}}.

\bibitem{Li:2020gnx}
H.-L. Li, Z.~Ren, J.~Shu, M.-L. Xiao, J.-H. Yu, and Y.-H. Zheng, ``{Complete
  set of dimension-eight operators in the standard model effective field
  theory},'' \href{http://dx.doi.org/10.1103/PhysRevD.104.015026}{{\em Phys.
  Rev. D} {\bfseries 104} no.~1, (2021) 015026},
  \href{http://arxiv.org/abs/2005.00008}{{\ttfamily arXiv:2005.00008
  [hep-ph]}}.

\bibitem{Li:2020xlh}
H.-L. Li, Z.~Ren, M.-L. Xiao, J.-H. Yu, and Y.-H. Zheng, ``{Complete set of
  dimension-nine operators in the standard model effective field theory},''
  \href{http://dx.doi.org/10.1103/PhysRevD.104.015025}{{\em Phys. Rev. D}
  {\bfseries 104} no.~1, (2021) 015025},
  \href{http://arxiv.org/abs/2007.07899}{{\ttfamily arXiv:2007.07899
  [hep-ph]}}.

\bibitem{Li:2020zfq}
H.-L. Li, J.~Shu, M.-L. Xiao, and J.-H. Yu, ``{Depicting the Landscape of
  Generic Effective Field Theories},''
  \href{http://arxiv.org/abs/2012.11615}{{\ttfamily arXiv:2012.11615
  [hep-ph]}}.

\bibitem{Li:2020tsi}
H.-L. Li, Z.~Ren, M.-L. Xiao, J.-H. Yu, and Y.-H. Zheng, ``{Low energy
  effective field theory operator basis at d \ensuremath{\leq} 9},''
  \href{http://dx.doi.org/10.1007/JHEP06(2021)138}{{\em JHEP} {\bfseries 06}
  (2021) 138}, \href{http://arxiv.org/abs/2012.09188}{{\ttfamily
  arXiv:2012.09188 [hep-ph]}}.

\bibitem{Murphy:2020rsh}
C.~W. Murphy, ``{Dimension-8 operators in the Standard Model Eective Field
  Theory},'' \href{http://dx.doi.org/10.1007/JHEP10(2020)174}{{\em JHEP}
  {\bfseries 10} (2020) 174}, \href{http://arxiv.org/abs/2005.00059}{{\ttfamily
  arXiv:2005.00059 [hep-ph]}}.

\bibitem{Li:2022tec}
H.-L. Li, Z.~Ren, M.-L. Xiao, J.-H. Yu, and Y.-H. Zheng, ``{Operators For
  Generic Effective Field Theory at any Dimension: On-shell Amplitude Basis
  Construction},'' \href{http://arxiv.org/abs/2201.04639}{{\ttfamily
  arXiv:2201.04639 [hep-ph]}}.

\bibitem{Sun:2022ssa}
H.~Sun, M.-L. Xiao, and J.-H. Yu, ``{Complete NLO Operators in the Higgs
  Effective Field Theory},'' \href{http://arxiv.org/abs/2206.07722}{{\ttfamily
  arXiv:2206.07722 [hep-ph]}}.

\bibitem{Elvang:2013cua}
H.~Elvang and Y.-t. Huang, ``{Scattering Amplitudes},''
  \href{http://arxiv.org/abs/1308.1697}{{\ttfamily arXiv:1308.1697 [hep-th]}}.

\bibitem{Henning:2019enq}
B.~Henning and T.~Melia, ``{Constructing effective field theories via their
  harmonics},'' \href{http://dx.doi.org/10.1103/PhysRevD.100.016015}{{\em Phys.
  Rev. D} {\bfseries 100} no.~1, (2019) 016015},
  \href{http://arxiv.org/abs/1902.06754}{{\ttfamily arXiv:1902.06754
  [hep-ph]}}.

\bibitem{ma2007group}
Z.~Ma, {\em Group Theory for Physicists}.
\newblock World Scientific, 2007.
\newblock \url{https://books.google.com.hk/books?id=E2hkWLiKjX4C}.

\bibitem{Kampf:2012fn}
K.~Kampf, J.~Novotny, and J.~Trnka, ``{Recursion relations for tree-level
  amplitudes in the $SU(N)$ nonlinear sigma model},''
  \href{http://dx.doi.org/10.1103/PhysRevD.87.081701}{{\em Phys. Rev. D}
  {\bfseries 87} no.~8, (2013) 081701},
  \href{http://arxiv.org/abs/1212.5224}{{\ttfamily arXiv:1212.5224 [hep-th]}}.

\bibitem{Li:2021tsq}
H.-L. Li, Z.~Ren, M.-L. Xiao, J.-H. Yu, and Y.-H. Zheng, ``{Operator bases in
  effective field theories with sterile neutrinos: d \ensuremath{\leq} 9},''
  \href{http://dx.doi.org/10.1007/JHEP11(2021)003}{{\em JHEP} {\bfseries 11}
  (2021) 003}, \href{http://arxiv.org/abs/2105.09329}{{\ttfamily
  arXiv:2105.09329 [hep-ph]}}.

\bibitem{Henning:2019mcv}
B.~Henning and T.~Melia, ``{Conformal-helicity duality \textbackslash{}\& the
  Hilbert space of free CFTs},''
  \href{http://arxiv.org/abs/1902.06747}{{\ttfamily arXiv:1902.06747
  [hep-th]}}.

\end{thebibliography}\endgroup

\bibliographystyle{utphys}


\end{document}